\documentclass[useAMS,usenatbib]{mn2e}
\usepackage{graphicx}

\usepackage{lipsum}
\usepackage{lscape}
\usepackage{longtable,amsmath}
\usepackage{booktabs}
\usepackage{marvosym}
\usepackage{pdflscape}
\usepackage[none]{hyphenat}
\usepackage[T1]{fontenc}
\usepackage{aecompl}
\usepackage{lscape}
\usepackage{capt-of}
\usepackage{multirow}
\usepackage{rotating}
\usepackage{array}
\usepackage{url}
\usepackage{verbatim}
\usepackage{float}
\usepackage{xcolor}

\newcommand{\mic}{\,$\umu$m}

\newcommand{\dg}{$^{\circ}$}

\newcommand{\htwo}{H$_2$}

\newcommand{\brg}{Br\,$\gamma$}
\newcommand{\ha}{H\,$\alpha$}

\newcommand{\hii}{H{\sc ii}}
\newcommand{\hei}{He{\sc i}}
\newcommand{\fdeg}{\,$^{\circ}$}

\newcolumntype{H}{>{\setbox0=\hbox\bgroup}c<{\egroup}@{}}

\title[Planetary Nebulae in UWISH2]{Planetary Nebulae in the UWISH2 survey}

\author[T.M.~Gledhill et
al.]{\parbox{\textwidth}{
T.M.~Gledhill$^{1}$\thanks{E-mail: t.gledhill@herts.ac.uk},
D.~Froebrich$^{2}$, J.~Campbell-White$^{2}$, A.M.~Jones$^{1}$
 \vspace{0.4cm}} \\
$^{1}$Centre for Astrophysics Research, University of Hertfordshire, College Lane, Hatfield AL10 9AB, UK \\ 
$^{2}$Centre for Astrophysics \& Planetary Science, The University of Kent, Canterbury, Kent CT2 7NH, UK \\  
}  

\begin{document}


\pagerange{\pageref{firstpage}--\pageref{lastpage}} \pubyear{2018}

\maketitle

\label{firstpage}

\begin{abstract}

Near-infrared imaging in the $1-0$\,S(1) emission line of molecular
hydrogen is able to detect planetary nebulae (PNe) that are hidden
from optical emission line surveys. We present images of 307 objects
from the UWISH2 survey of the northern Galactic Plane, and with the
aid of mid-infrared colour diagnostics draw up a list of
291 PN candidates. The majority, 183, are new
detections and 85 per cent of these are not present in \ha\ surveys of
the region. We find that more than half (54 per cent) of objects have
a bipolar morphology and that some objects previously considered as
elliptical or point-source in \ha\ imaging, appear bipolar in UWISH2
images. By considering a small subset of objects for which physical
radii are available from the \ha\ surface brightness-radius relation,
we find evidence that the \htwo\ surface brightness remains roughly
constant over a factor 20 range of radii from 0.03 to 0.6\,pc,
encompassing most of the visible lifetime of a PN. This leads to the
\ha\ surface brightness becoming comparable to that of \htwo\ at large
radius ($>0.5$\,pc). By combining the number of UWISH2 PNe without
\ha\ detection with an estimate of the PN detection efficiency in
\htwo\ emission, we estimate that PN numbers from \ha\ surveys may
underestimate the true PN number by a factor between 1.5 and 2.5
within the UWISH2 survey area.

\end{abstract}

\begin{keywords}  stars: evolution -- infrared: stars 
-- planetary nebulae: general -- ISM: individual: Galactic Plane
\end{keywords}

\section{Introduction}

There are currently around 3\,500 Galactic objects which are
considered to be true, likely, or possible planetary nebulae (PNe) and
these are now listed in the recently-developed HASH (Hong
Kong/AAO/Strasbourg/\ha) database (Parker, Boji\v{c}i\'{c} \& Frew
2016). The majority have been discovered at optical wavelengths by
their \ha\ emission and particularly from wide-field \ha\ surveys of
the Galactic plane, such as the SuperCOSMOS \ha\ Survey (SHS)(Parker
et al. 2005; Frew et al. 2014), the INT Photometric \ha\ Survey
(IPHAS) (Drew et al. 2005; Barentsen et al. 2014) and ongoing surveys
such as the VST Photometric \ha\ Survey (VPHAS+) (Drew et al. 2014).
Searches for PNe within the SHS survey have resulted in the
Macquarie/AAO/Strasbourg/\ha\ (MASH) catalogue (Parker et al. 2006;
Miszalski et al. 2008).  The first release of the IPHAS PN catalogue
is described by Sabin et al.\,(2014).  However, we also know that some
PNe will remain invisible to these optical surveys, due to obscuration
by dust, especially in the Galactic Plane (GP). In order to extend the
catalogued PN population to include these optically-obscured targets,
detection at longer wavelengths, particularly the near- and
mid-infrared (e.g. Cohen et al. 2011; Parker et al. 2012) will be
required.

PNe and pre-PNe (objects in the preceding post-AGB evolutionary stage
prior to ionization of the nebula) can be strong emitters of
\htwo\ emission lines, particularly the $1-0$\,S(1) line at
$\lambda=2.122$\mic\, in the near-infrared. After detection of
\htwo\ emission in NGC\,7027 (Treffers et al. 1976) a number of early
surveys reported \htwo\ emission in known PNe (Beckwith, Persson,
Gatley 1978; Storey 1984). Zuckerman \& Gatley (1988) established that
\htwo\ emission is more likely to be detected in PNe that are
morphologically classified as bipolar, that these objects tend to lie
at lower Galactic latitude, and by implication may derive from more
massive progenitors. A more extensive imaging survey of $\sim 100$ PNe
by Kastner et al. (1996) detected $1-0$\,S(1) emission in $\sim 40$
per cent of their targets. They confirmed the conclusions of Zuckerman
\& Gatley (1988), showing that 2/3 of bipolars in their sample were
detected in \htwo\ emission, compared with a detection rate of less
than 1 in 5 for non-bipolars. In a sample of 15 bipolar PNe, Guerrero
et al. (2000) find that objects with resolved equatorial ring
structures (R-BPNe) tend to be brighter in \htwo\ emission, and have
larger $1-0$\,S(1)/\brg\ flux ratios, than those with pinched waists
and unresolved cores (W-BPNe). These results are supported by more
recent work (Marquez-Lugo et al. 2015; Ramos-Larios et al. 2017).

The above-mentioned \htwo\ detections result from observation of
previously known PNe, identified in optical surveys.  The UKIRT
Wide-Field Imaging Survey for \htwo\ (UWISH2) is an unbiased
near-infrared imaging survey of the northern GP, conducted using the
WFCAM wide-field camera on the UK Infrared Telescope (UKIRT).  UWISH2
therefore presents the opportunity to search for new PNe candidates,
identified using their near-infrared \htwo\ emission.  The survey was
completed in 2013 and covers a contiguous area of the GP from
$l=357$\fdeg\ through the Galactic Centre to $l=65$\fdeg, with
latitudes generally within $b=\pm 1.5$\fdeg.  Additional smaller
fields include the star-forming regions in Cygnus and Auriga, giving a
total survey area of $\approx 287$\,deg$^2$.  A narrowband filter is
used, centred on the ${\rm v}=1-0$\,S(1) ro-vibrational line of
\htwo\ ($\lambda=2.122$\mic, $\delta\lambda=0.012$\mic).  The pixel
size is $0.2$ arcsec (with micro-stepping) with an exposure time of
$720$\,s per pixel. The median seeing over the whole survey is $0.8$
arcsec. The survey parameters are fully described in Froebrich et
al. (2011) and Froebrich et al. (2015), hereafter F15.

The resulting catalogue of extended emission includes more than
33\,000 individual features, and is described in F15.  Many of these
features are identified with star-forming activity, such as jets and
outflows (700 individual groups of features), or appear to be
associated with \hii\ regions or known supernova remnants.  F15 also
identified a total of 284 groups of features that are either
associated with known PNe or which are PN candidates, and these are
the subject of this paper. More than half of these objects are
previously unrecorded and not visible on SHS or IPHAS images,
revealing a potential population of optically-obscured PN and pre-PN
candidates.

\section{PN candidates}

\subsection{Candidate identification and imaging}

\begin{table*}
\caption{Summary table giving numbers of PN candidates in each morphological class.}
\begin{tabular}[l]{p{2cm}rrrp{1pt}rrrp{1pt}r}
\hline
           &  \multicolumn{3}{c}{\ha\ detected}&&\multicolumn{3}{c}{\ha\ not detected}&& \\
\cline{2-4} \cline{6-8}
           &  Known  & New  & N   && Known  & New & N  && Total \\
Bipolar    &  62     & 16   & 78  && 9      & 69  & 78 && 156   \\
Elliptical &  17     & 8    & 25  && 1      & 50  & 51 &&  76   \\
Round      &   6     & 2    &  8  &&        & 20  & 20 &&  28   \\
Asymmetric &   5     & 2    &  7  && 2      & 7   & 9  &&  16   \\
Irregular  &   5     &      &  5  && 1      & 8   & 9  &&  14   \\
Stellar    &         &      &     &&        & 1   & 1  &&   1   \\
\cline{4-4} \cline{8-8} \cline{10-10}
           &         &      &123  &&        &     &168 && 291   \\
\hline
\end{tabular}
\end{table*}

The method for identifying \htwo\ emission features in the survey
along with completeness estimation and rejection of contaminants is
described by F15 but briefly, the following steps were
followed: (i) the narrow-band images were continuum corrected by
subtracting a scaled $K$-band image taken from the UKIDSS GPS survey
(Lucas et al. 2008) to form a $H_2-K$ difference image; (ii) the
difference images were filtered to remove point sources, such as
bright stars, along with large-scale variations in the background;
(iii) an automated source detection algorithm was run to identify
extended \htwo\ features with a surface brightness greater than the
$5\,\sigma$ one-pixel noise and an area greater than 4 arcsec$^{2}$. 
The $5\,\sigma$ contour was used to define the outer bound
of the emission region and the number of pixels and flux within the
contour summed to give the total area on the sky and total flux of the
feature. Additionally, features lying close to bright stars, within
diffraction rings or within 10\,arcsec of an image edge, were rejected;
(iv) features separated by less than 3 arcmin were aggregated into
a `group' which is considered to be a single object.  These groups
were inspected to make sure that outliers around extended objects were
incorporated into that group and, conversely, that nearby but
obviously separate compact objects, were recorded as independent
groups.

\begin{figure}
\includegraphics[width=9.5cm,angle=0]{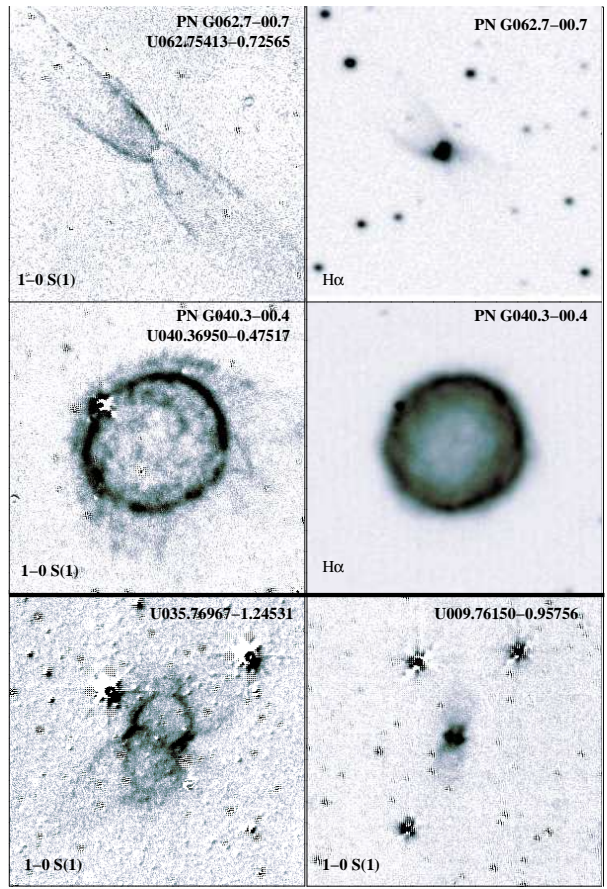}
\caption{Example UWISH2 images. The top and middle rows show two
  objects which are known to be PNe with $H_2-K$ difference images on
  the left and \ha\ images, from the IPHAS survey, on the right. The
  bottom row shows two new PN candidates which are not detected in
  \ha\ emission. The field of view in each case is 1 arcmin with North
  up and East left.}
\end{figure}

The resultant \htwo\ groups were examined by eye to assess whether or
not they may be PNe, based on morphology, proximity to star-forming
regions, association with known objects and near-infrared colour.
These steps resulted in a catalogue of 284 groups (F15) that are
either identified with known PNe or are considered to be PN
candidates; 261 objects are located in the main survey area, 16 in
Cygnus and 7 in Auriga. Fig.\,1 shows example $H_2-K$ difference
images for four objects; two are known PNe and two are new PN
candidates.  The ability of the $H_2-K$ images to detect faint
extended \htwo\ emission at high spatial resolution, while
simultaneously suppressing the many stellar point sources in the
field, is apparent. Similar images for all our UWISH2 PN candidates
can be found in Appendix\,B and C (available online).

In addition to the 284 objects mentioned above, we have searched for
nebulosity that may be associated with PNe amongst the
\htwo\ groupings originally classified as `u' by F15, and thought to
consist mostly of \hii\ regions. Of 1316 such groupings we identify a
further 22 objects as possible PN candidates, mainly on the basis of
their morphology, including one known PN, PN\,G030.2+01.5. In addition
we have reclassified 081.12354+1.23547 (MHO0362) from Young Stellar
Object (YSO) to possible PN. We list these extra PN candidates in
Table\,A2 in the Appendix. This gives a total of 307 extended
\htwo\ objects.

\subsection{Identification of non-PNe}
Cross-checking with the SIMBAD and HASH databases shows that 2 objects
are likely to be symbiotic systems and 3 are associated with
\hii\ regions. In addition, as described in Sec.\,5, we have used
mid-infrared colours from {\em Spitzer} IRAC photometry to try to
identify non-PNe, with the result that a further 5 objects are
classified as likely YSOs, 4 as \hii\ regions, 1 as a symbiotic system
and 1 as a likely pre-PN. A total of 16 objects are then eliminated
from the PN candidate list, leaving a total of 291.

\section{PN characteristics} 
All objects are listed in Table\,A1 with their UWISH2 identifier in
the form of the Galactic coordinates of the \htwo\ emission centre and
their status as PN or other object. We also assign a catalogue number
from 1 to 307 which is used to refer to objects more concisely within
this paper (e.g \#156).

\subsection{Morphology}
Objects have been classified morphologically according to their
appearance in the $H_2-K$ images.  We use the `ERBIAS' scheme (Parker
et al. 2006) which splits objects into Elliptical, Round (aspect ratio
less than 5 per cent), Bipolar, Irregular, Asymmetric and Stellar
(point source) groupings. A system of sub-classifiers, `amprs' can be
used to flag asymmetric, multiple, point-symmetric, ring-like and
resolved (internal) structures.

A summary of the classification into basic morphological types is
given in Table\,1. We note those objects that have been flagged as PNe
or possible PNe in the literature (Known), and those which are
unrecorded prior to UWISH2 (New). The majority of objects are bipolar
156 or 54 per cent) with 85 of these being new detections. The
remaining objects are mostly in the elliptical or round category (104
or 36 per cent in total) and of these 77 per cent are new
detections. There are relatively few objects in the remaining
asymmetric, irregular or stellar categories (31 or 11 per cent in
total).

\subsection{H$\alpha$ emission}

The UWISH2 area overlaps with the SHS (Parker et al. 2005) in the
Galactic Centre and inner GP regions out to $l\approx 35$\fdeg, and
with the IPHAS (Drew et al. 2005) for $l>29$\fdeg, so that any point
in UWISH2 is contained within one or both of these \ha\ surveys. We
have searched for \ha\ emission for each of our \htwo\ PN candidates
and, if found, extracted the image to compare emission extent and
morphologies. In a few cases (e.g. for fields which overlap with SHS)
better quality \ha\ images were available from other surveys such as
the ``IAC Morphological Catalog of Northern Galactic Planetary
Nebulae'' (Manchado et al. 1996) or from {\em HST}
imaging. \ha\ images are shown alongside \htwo\ images for individual
objects in Appendix\,B.

Objects are divided into two main groups: those that are detected in
\ha\ surveys and those that are not detected in these surveys, as
shown in Table\,1. Of the 291 \htwo\ emission-line objects identified
as PNe candidates in the UWISH2 survey, the majority (183) are not
previously recorded and more than 85 per cent of these (155) are not
detected in \ha. This is perhaps not surprising, given that the
principal discovery technique for PNe has been identification in
\ha\ surveys. However, we also find 28 objects for which \ha\ emission
is detected, but which have not been noted previously in the
literature.

\begin{figure}
\includegraphics[width=8.5cm,angle=0]{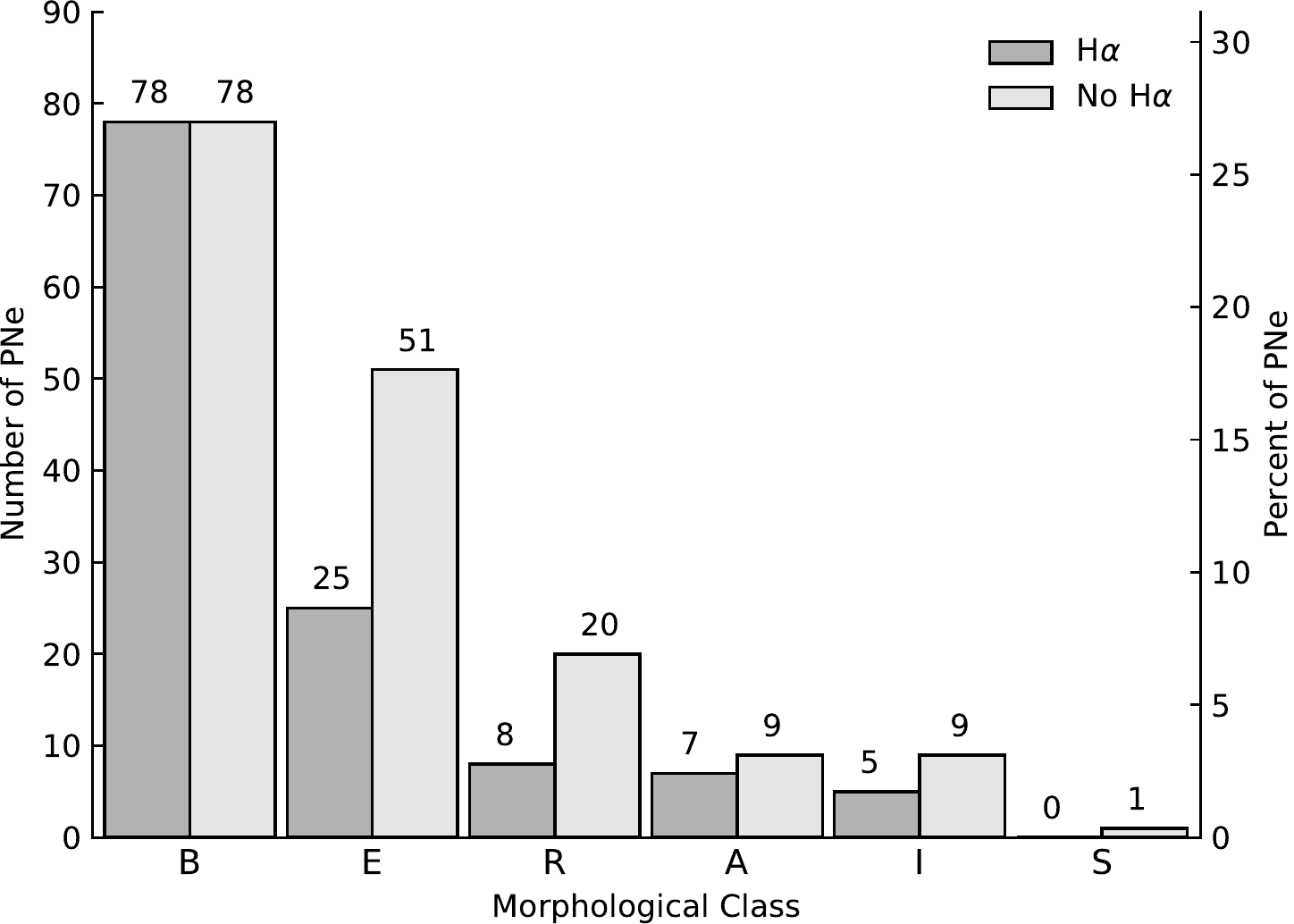}
\caption{Graphical representation of the number of H$_2$-detected PN
  candidates in each morphological class, divided between those
  appearing in (darker bars) and absent from (lighter bars) H$\alpha$
  surveys, with the number in each group at the top of the bar.}
\end{figure}

In Fig.\,2 we show graphically the number of H$_2$-detected PN
candidates, in each morphological category and whether they are
visible in IPHAS and/or SHS \ha\ surveys. Half of the objects in the
bipolar category with \htwo\ emission also have a \ha\ detection,
whereas in the elliptical, round and other categories the majority of
our \htwo-detected PNe candidates are not detected in the
\ha\ surveys.

Morphological classification is clearly subjective to some degree, but
where objects are detected in both UWISH2 and \ha\ surveys, we can
compare their morphological classification in the two emission
lines. This is summarised in Table\,2. We see that of the 50 objects
classed as bipolar (B) according to their \ha\ emission, 44 (or 88 per
cent) also appear bipolar in their \htwo\ emission, confirming that
\htwo\ imaging provides a comparable diagnostic of PN structure in
these cases. Conversely, of the 77 objects classed as bipolar in
\htwo, 44 (57 per cent) are bipolar in \ha, with 20 (26 per cent) and
8 (10 per cent) classed as elliptical and stellar, respectively. There
are two likely reasons for this: (i) recombination emission often
originates from the central regions of PNe, especially in young
objects, so may not reflect the larger-scale structure of the PN. For
example, the \ha\ emission may highlight the central waist of a larger
bipolar outflow seen in molecular emission; (ii) the spatial
resolution of the UWISH2 survey (median seeing 0.8 arcsec) is greater
than that of the IPHAS, and especially the SHS, so that structure seen
in \htwo\ may remain unresolved in \ha\ images in the case of compact
objects.

\subsection{Fraction of known PNe detected in UWISH2}
To assess the fraction of previously recorded PNe that are detected in
\htwo, we select objects from the HASH PN database (Parker,
Boji\^{c}i\'{c} \& Frew 2016) in the region $66$\fdeg\ $> l >
10$\fdeg\ and $|b|<1.5$\fdeg, where there is a fairly uniform coverage
in UWISH2. We consider only objects classed as `True' (T), `Likely'
(L) or `Possible' (P) PNe, discarding other non-PN classes, leaving
287 objects, of which 114 are T, 48 are L and 125 are P. The number of
objects with a positive \htwo\ detection is 85, with 61 (54 per cent),
15 (31 per cent) and 9 (7 per cent) for T, L and P respectively. More
than half of the spectroscopically confirmed (T) PNe in this region of
the GP have \htwo\ emission. Given the low fraction of
\htwo\ detection (7 per cent) in the case of `Possible' PNe, it seems
likely that a large fraction of these objects are not PNe.

\begin{figure}
\includegraphics[width=9cm,angle=0]{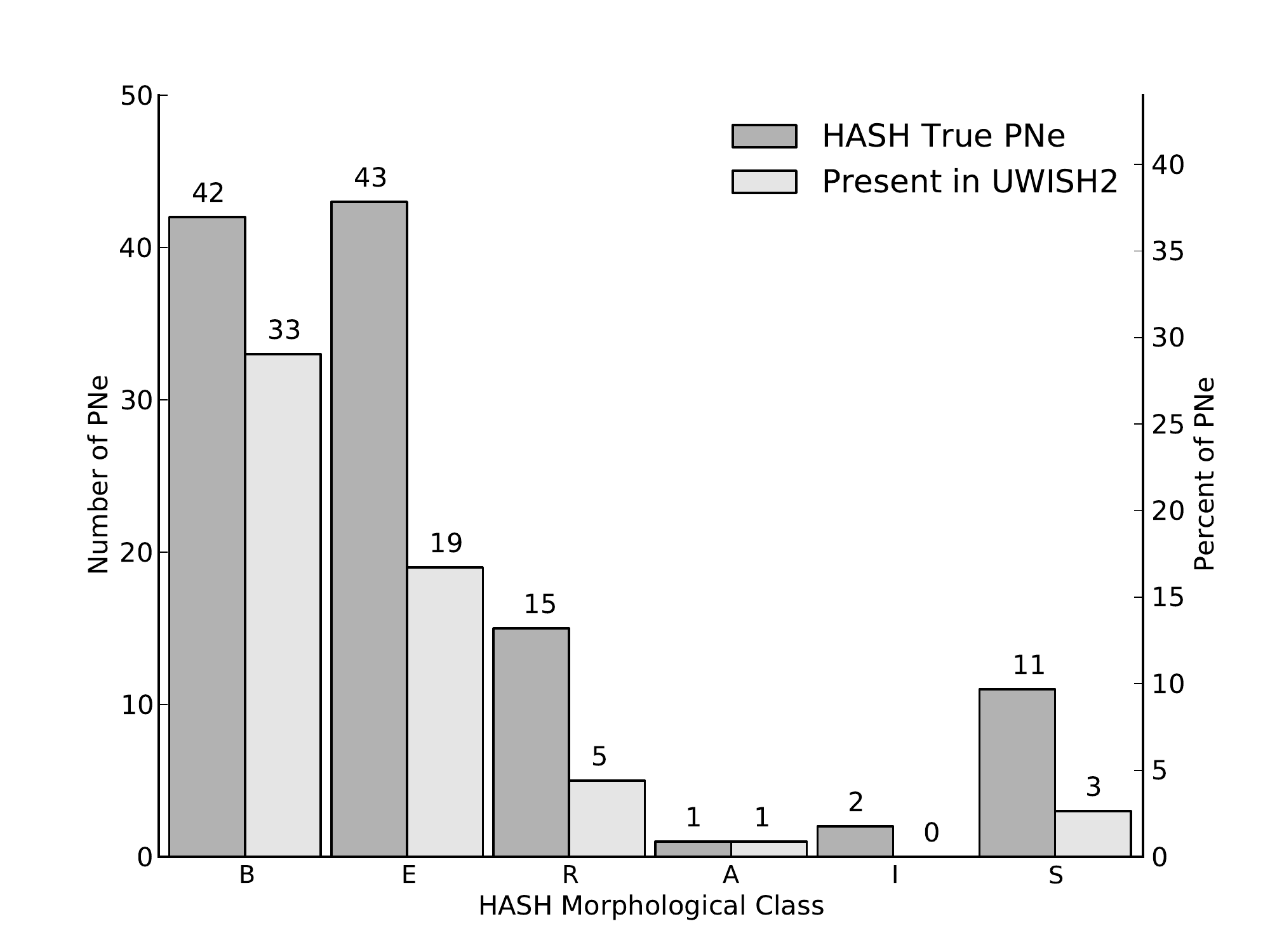}
\caption{The sample of `True' PNe from the HASH database in the region
  $66\degr > l > 10\degr$ and $|b|<1.5\degr$, grouped according to
  their morphological class as given in HASH (dark grey). The number
  of PNe in each class with a positive detection in UWISH2 is shown in
  light grey.}
\end{figure}

We group the T PNe according to their morphological class as given in
the HASH database, which is mostly determined by their appearance in
\ha\ imaging. We find that 33 out of 42 bipolar PNe are found in
UWISH2 (79 per cent), with a lower fraction of detections amongst E, R
and S classes (Fig.~2). 

Given that only 9 of the HASH bipolar `True' PNe in the selected
region are absent from UWISH2, it is worth looking at these more
closely. Two objects would not have been detected; PN\,G032.9-01.4
lies just off a UWISH2 tile, so is not included in the survey, whereas
PN\,G026.9-00.7 is located too close to a bright star and associated
diffraction pattern. A further two objects, PN\,G020.9-01.1 (M1-51)
and PN\,G029.2+00.0 appear in the UWISH2 narrow-band $H_2$ images, but
are negative in their $H_2$--$K$ difference image so have not been
flagged as detections.  This is most likely due to strong
\brg\ emission in the $K$-band, so that we cannot say for sure if
\htwo\ emission is present or not.  The remaining 5 objects
(PN\,G026.0-01.1, PN\,G027.4-00.9, PN\,G042.7+00.8, PN\,G045.7+01.4
and PN\,G063.5+00.0) are not detected, so the fraction of HASH True PN
with bipolar morphology detected in UWISH2 could be as high as 37/42,
or 88 per cent.

There is a subtlety here though, in that the \ha\ morphology does not
map directly onto the \htwo\ morphology. As mentioned, a significant
number of objects with E or S morphologies as determined from their
\ha\ emission, appear bipolar in \htwo\ emission (Table\,2) when
observed with higher spatial resolution.  This could have the effect
of increasing the number of bipolar PN that are not detected in
UWISH2.  It has also been noted that \htwo\ morphology often
correlates more closely with structure seen in [O\,{\scriptsize I}]
630.0\,nm (Beckwith et al. 1978) and [N\,{\scriptsize II}] 658.3\,nm
(Fang et al. 2018), than with \ha.

\setlength{\extrarowheight}{1mm}
\begin{table}
\caption{A comparison of morphological classification for objects detected in 
both UWISH2 and \ha\ surveys.}
\begin{tabular}{p{2mm}p{2mm}cccccc}
          && \multicolumn{6}{c}{\ha\ classification} \\
          && B & E & R & A & I & S \\
          \cline{3-8} 
\end{tabular}
\begin{tabular}{p{2mm}p{2mm}|cccccc}
\multirow{6}{*}{\begin{turn}{-90}{\htwo\ classification} \end{turn}} 
          &B&44&20&1&2&2&8 \\
          &E&2&14&8&1&0&0 \\
          &R&2&2&4&0&0&0 \\
          &A&1&2&1&2&0&1 \\
          &I&1&1&1&1&1&0 \\
          &S&0&0&0&0&0&0 \\
\end{tabular}
\end{table}

\subsection{Spatial distribution}

The spatial distribution of \htwo-detected PN candidates is shown in
the upper diagram of Fig.\,4.
\begin{figure*}
\includegraphics[width=19cm,angle=0]{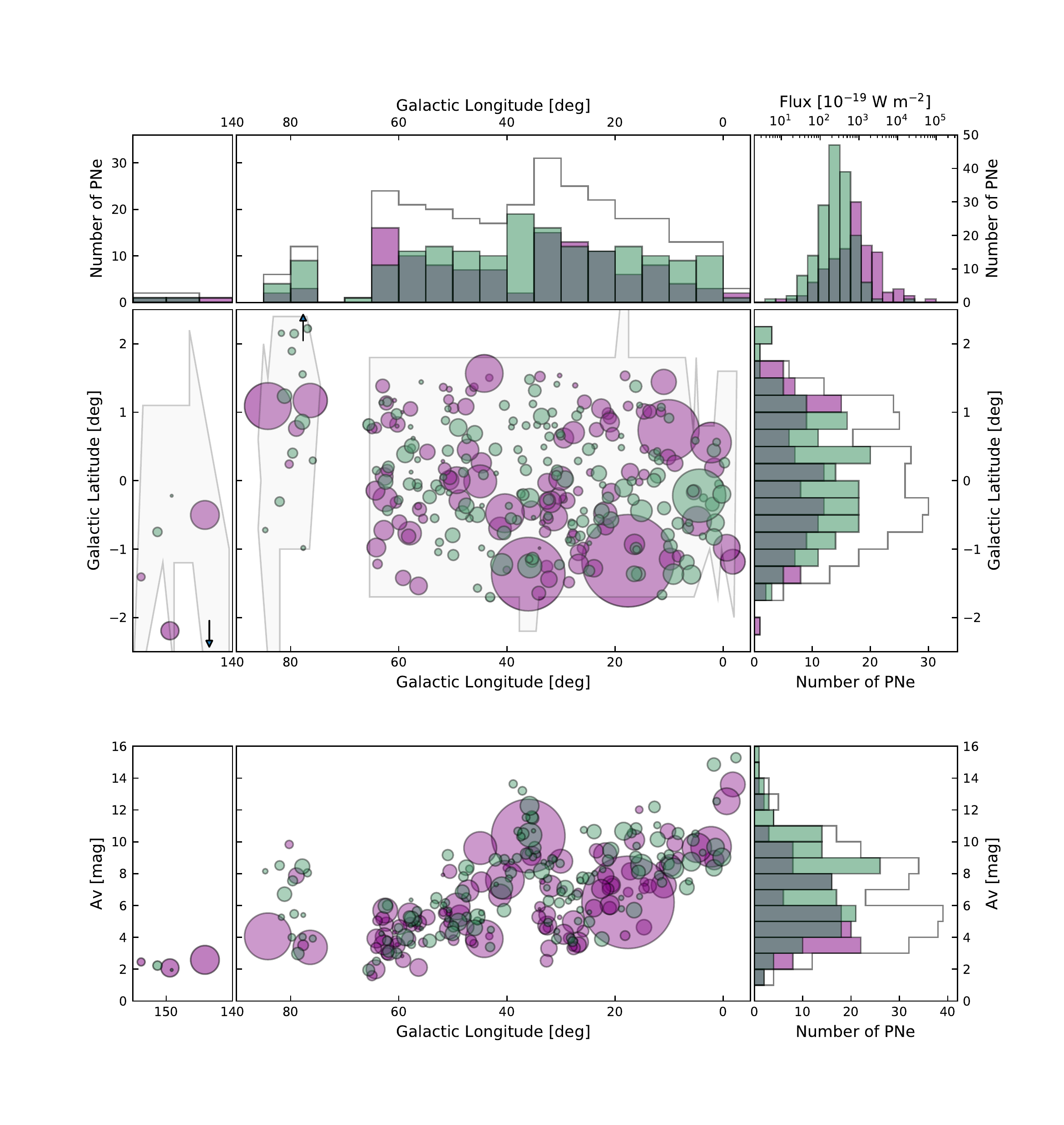}
\caption{The top portion of the diagram shows the distribution of
  H$_2$-detected PN candidates in Galactic coordinates. Objects detected in
  \ha\ emission are coloured purple and those not detected, green. The
  circle diameter is proportional to the area of \htwo\ emission on the sky
  (given in F15). The main
  survey area is shown, along with the smaller fields in Cygnus and
  Auriga, with grey polygonal outlines indicating the survey
  footprint. Arrows indicate two objects lying off the plot in
  latitude.  Histograms of the longitude and latitude distribution
  also show the total PN numbers as a stepped black line. The flux
  distribution is shown in the top right. The lower diagram shows an
  estimate of the extinction at the sky position of each object as a
  function of Galactic longitude, with the distribution to the right.}
\end{figure*}
The general distribution was discussed by F15 who noted that while a
KS test gives a probability of 96 per cent that the distribution of PN
in Galactic longitude is homogeneous, there are approximately 10 per
cent fewer PN per unit area in the inner GP ($l<30\degr$) than the
outer GP, suggesting that this may be due to increased extinction
towards these sightlines. The Galactic latitude distribution is more
difficult to determine due to the limited coverage of UWISH2 in this
direction but a scale height of 0\fdeg.92\,$\pm$\,0\fdeg.11 was
determined by F15, with the zero point coincident with $b=0.0$ within
uncertainties. The distribution of total PN number with $l$ and $b$ is
shown in Fig.\,4 by the black stepped lines in the longitude and
latitude histogram panels. The total PNe candidate number per
longitude bin increases gradually from the Galactic centre outwards,
peaking at 30 objects in the range $l=30-35$\degr, before dropping
sharply to below 20 and then resuming a gradual increase.

The breakdown into objects with and without \ha\ detection is shown by
purple and green circles respectively, with the circle size indicative
of the area of \htwo\ emission on the sky. There are 168 objects with
no \ha\ detection (green), or 58 per cent of the total, with many seen
to be small (and faint) relative to objects detected in \ha. The
median surface area of \htwo\ emission is 162 and 70 arcsec$^{2}$ for
objects with and without \ha\ detection. The distribution in Galactic
longitude of objects with \ha\ detection is noticeably different to
that of objects without; there are relatively fewer PN candidates with
\ha\ detection in the inner GP ($l < 20\degr$) and a pronounced dip in
number between 35\degr and 40\degr.  The distribution with latitude
shows generally more objects without \ha\ detection at low latitudes
($-1\degr < b < 1\degr$) and the reverse trend outside this region.

\subsubsection{Extinction Effects}
To investigate the effect of dust obscuration on the PN spatial
distributions, we have used the extinction maps of Rowles \& Froebrich
(2009). These maps are based on the median near-infrared colour excess
of the nearest 49 stars to a given sky position, calculated from 2MASS
photometry. This is converted to an estimate of $A_{\rm v}$ on a grid
with spatial resolution of approximately 1 arcmin in the GP. As
explained by Rowles \& Froebrich (2009), the use of the median colour
excess (rather than the mean) helps to reduce bias in the extinction
estimate due to intrinsically red young stars, or cluster stars with
colours different to foreground/background stars. However, this also
means that if there are few background stars along a line of sight,
then the foreground stars dominate and the extinction is
underestimated. This can easily happen in highly extincted regions of
the GP where there are few background stars, even in the
near-infrared. To mitigate against this, we have examined the
extinction map to identify discontinuous regions where the extinction
jumps to lower values over a resolution element and have discounted 24
objects that lie in these regions. The remaining 267 objects are
plotted in the lower diagram of Fig.\,4.

As expected there is a general trend of falling $A_{\rm v}$ with
increasing $l$, from $\sim 15$~mag toward the Galactic Centre to $\sim
2$~mag in the Auriga field. There are upward spikes of extinction in
the Cygnus clouds and in the region $35\degr < l< 40\degr$,
corresponding to dark clouds in the Aquilla region. The latter region
in particular coincides with the dip in \ha -detected PN noted above,
which is therefore most likely caused by increased extinction along
these lines of sight.  The distribution of PN number with $A_{\rm v}$
(lower, right panel in Fig.\,4) shows relatively fewer PNe with
\ha\ detections in regions characterised by higher extinction ($A_{\rm
  v} > 5$); the median extinction for regions containing PN with and
without \ha\ detection is 5.4 and 7.2 mag, respectively.

\subsection{Flux and Surface Brightness Distribution}
The background-corrected \htwo\ emission in each object can be summed
over all regions lying above the $5\sigma$ pixel noise contour in the
$H_{2} - K$ difference image, to give a total flux, $F_{\rm
  tot}$. However this method can lead to an over- or under-estimate in
cases where the $5\sigma$ contour includes positive or negative
residuals, usually due to imperfect bright star subtraction. This can
be a particular problem for very extended objects. An alternative
method is to sum the median flux over the object (i.e. the median flux
of a region multiplied by its area) to give $F_{\rm med}$. This will
be less biased by point source subtraction errors, but is likely to
underestimate the flux in objects where most of the \htwo\ emission
arises from compact regions, such as knots and rims, which are common
in our targets. Hence both methods have their drawbacks (see F15 for a
more detailed comparison).

\begin{figure}
\includegraphics[width=9cm,angle=0]{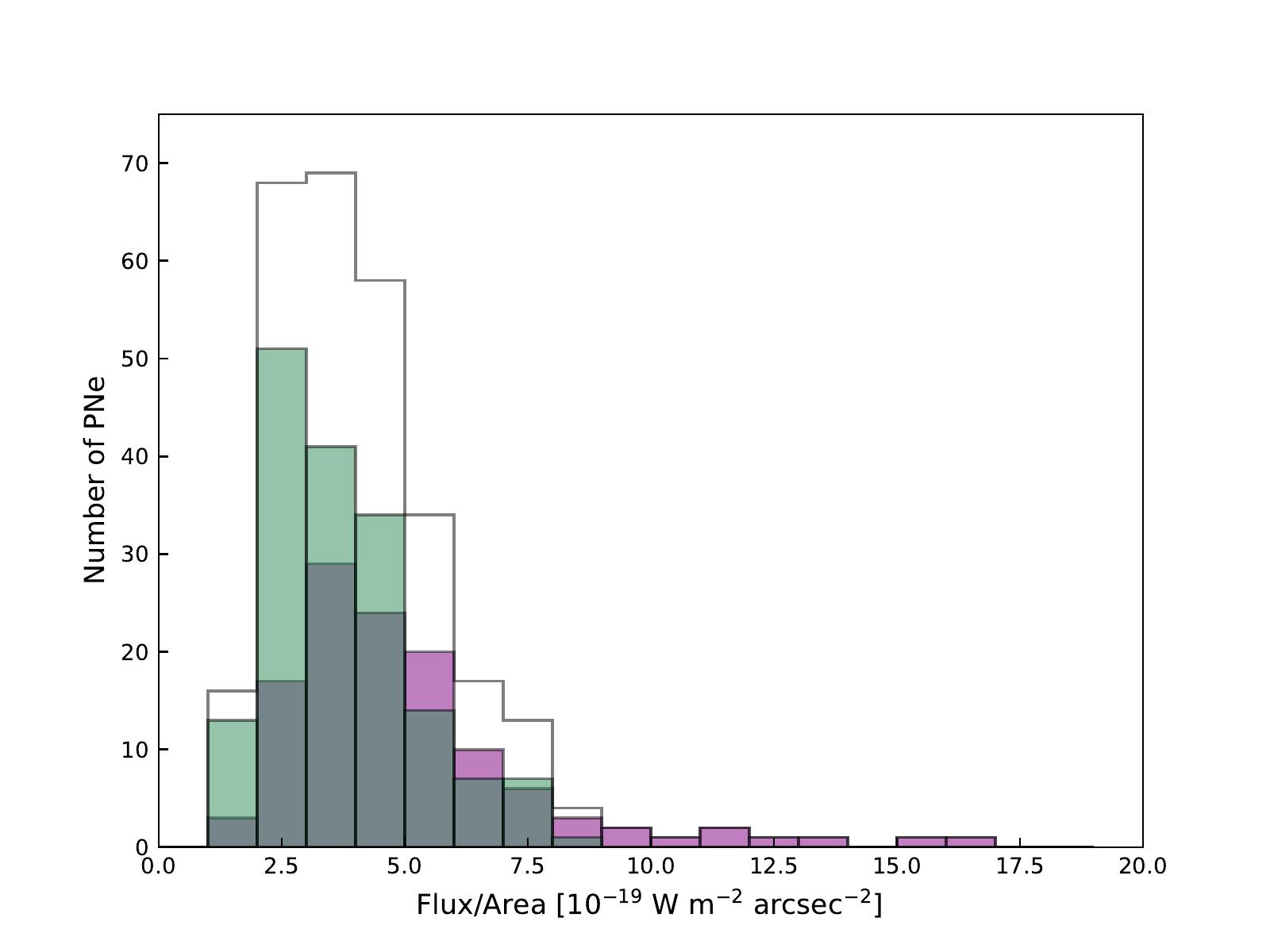}
\caption{H$_2$ surface brightness distribution for objects with
  (purple) and without (green) \ha\ detection. The total number is
  shown as a stepped black line.  The $F_{\rm med}$ flux estimator is
  used to reduce the sparse tail of objects at high flux due to
  residual star subtractions (see text).}
\end{figure}

The $F_{\rm tot}$ flux distributions for objects with and without
detectable \ha\ emission are shown, on a log scale, in the top right
panel of Fig.\,4. The two distributions are clearly different, with a
pronounced peak due mostly to fainter objects not detected in \ha ,
and with the peak of the \ha -detected sample lying at higher
flux. The bright tail of the flux distribution will include some
objects with over-estimated flux due to stellar residuals as described
above. The median values for the flux distributions with and without
detected \ha\ emission are $9.1\times 10^{-17}$ and $3.2\times
10^{-17}$~W~m$^{-2}$, respectively, independently of whether $F_{\rm
  tot}$ or $F_{\rm med}$ is adopted. This strongly suggests that
UWISH2 is detecting a population of new and faint objects which are
not visible in \ha\ emission and which, in combination with the larger
extinction towards these objects, are likely to be optically-obscured
PN candidates.

The surface brightness (i.e. \htwo\ flux divided by area of emission
on the sky) is, however, very similar across the UWISH2 sample,
regardless of \ha\ detection, and is shown in Fig.\,5. The median
surface brightness is $3.9\times 10^{-19}$~W~m$^{-2}$~arcsec$^{-2}$,
and is again independent of whether the $F_{\rm tot}$ or $F_{\rm med}$
flux estimator is adopted. The narrow distribution in surface
brightness (neglecting a few bright objects detected in \ha) means
that fainter objects tend also to have smaller angular extent, and
suggests that the difference in the median flux of the objects with
and without \ha\ detection (Fig.\,4) is due to the latter being, on
average, at larger distance. This is again consistent with the idea
that the near-infrared UWISH2 survey is probing longer sight lines to
more distant and reddened objects.

\section{Surface brightness-radius relationships}

The relationship between the \ha\ surface brightness and physical
radius of an ionized nebula has been used to derive distances to more
than 1000 PNe (Frew, Parker \& Boji\v{c}i\'{c} 2016, hereafter FPB16).
The relationship arises due to the \ha\ flux scaling with the nebula
volume (i.e. with the cube of the radius, $r$) and the emissivity,
which for recombination line emission depends on the square of the
electron density, $n_{\rm e}$. The surface brightness, $S$, is
independent of distance and scales as $rn_{\rm e}^2$. Assuming a power
law $n_{\rm e} \propto r^{\alpha}$ then $S \propto r^{2\alpha +1}$.
FPB16 find $S_{\rm H\alpha}\propto r^{-3.6}$ for a sample of
calibration PNe for which the distances (and hence radii) are
independently known.

There are 23 PNe for which FPB16 derive radii and distances based on
their extinction-corrected \ha\ surface brightness, and which also
appear in UWISH2 (Table~3). $S_{\rm H\alpha}$ for these objects is
plotted in Fig.\,6 as green circles, and by definition they lie on a
straight line of slope $-3.6$ in a log-log plot, as their physical
radii have been estimated assuming the above \ha\ surface
brightness-radius ($S_{\rm H\alpha}-r$) relationship.  The 38 objects
listed in FPB16 which are within the UWISH2 survey area, but not
detected in \htwo, are shown as black `plus' symbols and listed at the
foot of Table\,3. The dotted horizontal line shows an indicative
\htwo\ surface brightness detection limit of $1.62\times
10^{-19}$~W~m$^{-2}$~arcsec$^{-2}$, based on the median pixel noise
for the UWISH2 survey (note that the actual pixel noise varies across
the survey; see F15 and their fig.\,4 for details).  The red circles
show the $1-0$~S(1) surface brightness (i.e. $F_{\rm med}$ divided by
area of \htwo\ emission) for the detected objects. In order to correct
the \htwo\ data for extinction, for comparison with the de-reddened
\ha\ data, we use the same $E(B-V)$ colour excess given by FPB16. We
assume a ratio of total to selective extinction $A_{\rm V}/E(B-V)=3.1$
and a power-law near-infrared extinction curve with index $-1.95$
(Wang \& Jiang 2014).

The \htwo\ surface brightness-radius ($S_{\rm H_2}-r$) relation,
unlike the \ha\ emission, appears roughly flat over a factor of 20 in
nebula radius, from 0.03 to 0.6~pc. The best-fit line shown has a
gradient of $-0.14$. This range of radii is expected to encompass most
of the visible (optical emission line) lifetime of a PN (Jacob et
al. 2013), ranging from young, bright, and compact objects on the left
to extended, faint and old objects to the right. Note that the radius
here is that of the ionised region and the \htwo\ emission may extend
out to larger radius in the form of a molecular envelope. The objects
with smallest and largest radius, \#214 and \#252, have low surface
brightness compared to the average trend. Object \#214 has intense
\brg\ emission from a compact region in the centre; this leads to
over-subtraction in the $H_2-K$ difference image and a likely
under-estimate of the \htwo\ flux, so that this point is a lower limit
to the true surface brightness. In the case of \#252, the
\htwo\ emission is genuinely very faint, possibly indicating that in
the most extended and presumably most evolved objects, little
molecular material remains. Neglecting these two outliers, the mean
\htwo\ surface brightness is $8.6 \pm 3.6 \times
10^{-19}$~W~m$^{-2}$~arcsec$^{-2}$ over a factor of 10 range in PN
radius, from 0.05~pc to 0.5~pc.

\begin{figure}
\includegraphics[width=8.5cm,angle=0]{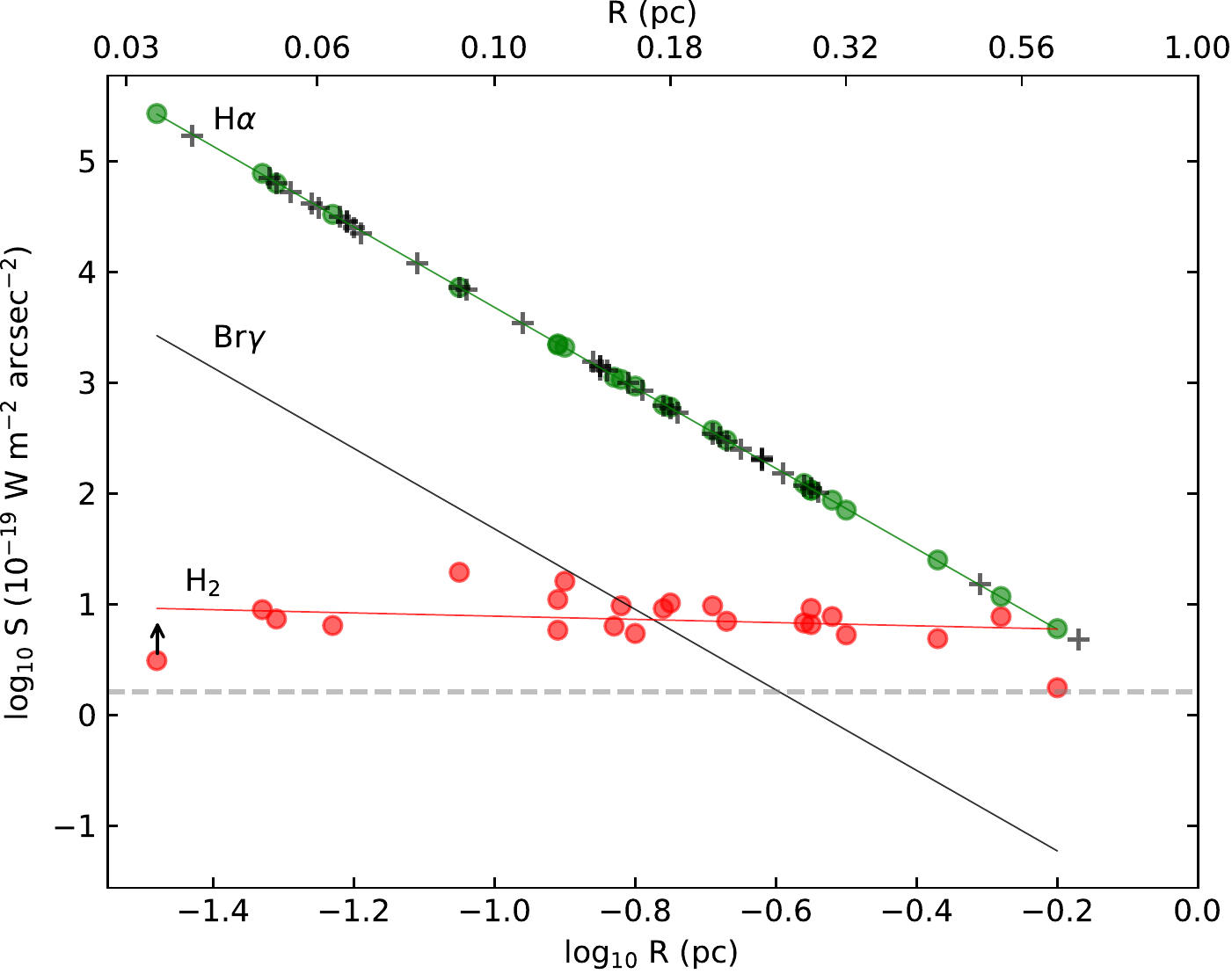}
\caption{Surface brightness against radius for 23
  PNe in common between UWISH2 and FPB16 with green
  circles showing \ha\ emission and red
  circles $1-0$~S(1) emission. Lines of best fit are
  also shown. The \brg\ trend is obtained by shifting the \ha\ fit
  line down by 2 dex.  PNe from FPB16 which are
    within the UWISH2 area but not detected in \htwo\ are shown as
    black `plus' symbols. The \htwo\ surface brightness detection
    limit is shown as a dotted line (see text for details).}
\end{figure}

\renewcommand{\arraystretch}{0.8}
\begin{table}
\caption{23 UWISH2 PNe with radii (in increasing order) taken from
  FPB16.  The ratio of \brg\ to \htwo\ surface brightness is estimated
  by scaling the \ha\ surface brightness, as explained in the
  text. The value for \#214 is an upper limit. Median \htwo\ flux,
  $F_{\rm med}$, is in units of $10^{-19}$\,W\,m$^{-2}$.  The PN\,G
  numbers of 38 objects in FPB16 which are within the UWISH2 area but
  not detected in \htwo\ are shown at the bottom of the table.}
\begin{tabular}{lHHlllHl} \\
\hline
Object \# &PN G   & UWISH2 ID & Other Name & $\log_{10} r$ & S$_{\rm{Br}\gamma}$/S$_{\rm{H_2}}$ & Morph. & $F_{\rm med}$   \\
\hline
214&055.5-00.5 &055.50747-0.55729 &M~1-71     & -1.48             & 1.24e3$\downarrow$   & E &139$\uparrow$  \\
026&010.1+00.7 &010.10373+0.73752 &NGC~6537   & -1.33             & 1.06e2   & B  &20\,474 \\
021&008.3-01.1 &008.33574-1.10291 &M~1-40     & -1.31             & 7.17e1   & I  &552 \\
098&027.7+00.7 &027.70327+0.70354 &M~2-45     & -1.23             & 8.04e1   & R  &1\,557 \\
284&359.3-00.9 &359.35683-0.98000 &Hb~5       & -1.05             & 2.73e0   & B  &8\,645\\
087&025.9-00.9 &025.92671-0.98449 &Pe~1-14    & -0.91             & 2.37e0   & B  &858\\
262&077.7+03.1 &077.68068+3.12797 &KjPn~2     & -0.91             & 4.36e0   & B  &414 \\
074&022.5+01.0 &022.57000+1.05505 &MaC~1-13   & -0.90             & 1.69e0   & B  &3\,279 \\  
072&021.8-00.4 &021.81951-0.47837 &M~3-28     & -0.83             & 2.23e0   & B  &2\,075\\
071&021.7-00.6 &021.74338-0.67287 &M~3-55     & -0.82             & 1.31e0   & B  &1\,480\\
280&153.7-01.4 &153.77044-1.40652 &K~3-65     & -0.80             & 2.16e0   & B  &202\\
131&032.9-00.7 &032.94004-0.74662 &CBSS~3     & -0.76             & 8.92e-1   & B &698\\
161&040.3-00.4 &040.36950-0.47517 &Abell~53   & -0.75             & 6.52e-1   & R &9\,475\\
128&032.5-00.3 &032.54998-0.29529 &Te~7       & -0.69             & 5.23e-1   & B &1\,552\\
118&031.3-00.5 &031.32618-0.53286 &HaTr~10    & -0.67             & 5.26e-1   & B &3\,019\\
044&014.6+01.0 &014.65833+1.01220 &PHR~J1813-1543 & -0.56         & 2.07e-1   & E &882\\
259&076.3+01.1 &076.37264+1.17216 &Abell~69   & -0.55             & 1.44e-1   & B &4\,463\\
272&084.2+01.0 &084.20031+1.09069 &K~4-55     & -0.55             & 2.17e-1   & B &12\,704 \\
120&031.9-00.3 &031.90685-0.30936 &WeSb~4     & -0.52             & 1.40e-1   & B &2\,182\\
147&035.9-01.1 &036.05309-1.36593 &Sh~2-71    & -0.50             & 1.45e-1   & B &9\,476\\
047&015.5-00.0 &015.53753-0.01923 &PHR~J1818-1526 & -0.37         & 5.97e-2   & E &178\\
165&041.2-00.6 &041.27043-0.69797 &HaTr~14    & -0.28             & 1.55e-2   & B &2\,883 \\
252&063.9-01.2 &063.92454-1.21740 &Te~1       & -0.20             & 3.73e-2   & B &108\\
\hline 
\multicolumn{8}{p{8cm}}{000.1-01.7, 000.2+01.7, 000.3-01.6,
  000.6-01.3, 000.7-01.5, 000.8+01.3, 000.8-01.5,
  000.9+01.1, 000.9-01.2, 001.0+01.3, 001.1-01.6, 001.2-01.2a,
  001.3-01.2, 010.0-01.5, 011.7+00.0, 011.7-00.6, 013.3+01.1,
  015.5+01.0, 017.5+01.0, 019.6+00.7, 019.9+00.9, 020.2-0.06,
  020.9-01.1, 027.0+01.5, 027.5+01.0, 029.0+00.4, 036.9-01.1,
  056.4-00.9, 058.9+01.3, 060.5-00.3, 065.9+00.5, 151.4+00.5,
  358.9-00.7, 359.1-01.7, 359.2+01.3, 359.3-01.8, 359.5-01.2,
  359.7-01.4}\\
\hline
\end{tabular}
\end{table}  
\renewcommand{\arraystretch}{1.0}

The similarity of \htwo\ surface brightness in objects covering such a
large range of radii suggests that, unlike \ha, the \htwo\ emission
does not originate throughout the volume of the PN and does not scale
in the same way with radius.  This is consistent with many of our
UWISH2 images (Appendix B and C) which show that \htwo\ emission often
arises from thin structures at the extremities of PNe, such as cavity
walls and edges in bipolar objects, and rims and shells in more
elliptical objects.  This has been noted in previous studies
(e.g. Guerrero et al.  2000; Fang et al. 2018) and is particularly
evident in \#161 (PN\,G040.3-00.4; Abell\,53), shown in Fig.\,1, where
the \htwo\ emission arises mainly from a thin circular ring. Assuming
that the object is a spherically symmetric PN, rather than a bipolar
viewed exactly pole-on, then this corresponds to a thin shell at the
outer boundary of the ionised region, the extent of which is indicated
by the \ha\ image. This is also consistent with, for example, the
spherically-symmetric photoionization models of Aleman \& Gruenwald
(2011) which show that the most significant contribution to $1-0$~S(1)
emission occurs in a relatively thin transition zone (TZ) outside the
fully ionised region. They define the TZ to be a shell of warm,
partially ionised gas, where the fraction of ionized hydrogen drops
below 0.95 and find that the $1-0$~S(1) emissivity, $j$, rises to a
maximum in this region.  We note, however, that these models do not
include a molecular envelope outside the ionized region, or shock
excitation of the \htwo\ molecule, which is known to be important,
especially in bipolar objects (e.g. Kastner et al.  1996; Marquez-Lugo
et al. 2015). In fact, there is mounting evidence that the relative
importance of shocks and fluorescence depends on PN age, with shocks
being prevalent in early and late phases of PN evolution (Davis et
al. 2003; Ramos-Larios et al. 2017).

Fig.\, 6 shows that for PNe with large radii ($r > 0.5$~pc) the
\htwo\ and \ha\ surface brightnesses become comparable. If the angular
extents in both lines are also similar, a reasonable assumption for
evolved objects where the ionisation front will have overtaken most of
the molecular envelope, then the \htwo\ and \ha\ fluxes should also be
similar. In these cases, detection of extended PNe using their
\htwo\ emission may be easier, especially considering a factor of
$\sim 10$ in differential extinction between the two lines. 

We also plot the approximate \brg\ surface brightness-radius relation
in Fig.~5, by shifting the \ha\ surface brightness from FPB16 (already
dereddened) down by 2 dex to account for the intrinsic \brg /
\ha\ line ratio\footnote{The value of $\log({\rm Br}\gamma /{\rm
    H}\alpha )$ varies weakly with gas temperature and density but is
  $-2.0\pm0.1$ for values in the range $10^{3}-3\times10^{4}$~K and
  $10^{2}-10^{4}$~cm$^{-3}$ respectively (Hummer \& Storey
  1987).}. Due to the flat trend in \htwo\ surface brightness, the
ratio of \brg\ to \htwo\ surface brightness also varies with nebula
radius, with the two being comparable for a radius of approximately
0.17~pc, as shown by the intersection of the two best-fit lines
(Fig.\,6). Smaller PNe have $S_{\rm{Br}\gamma} > S_{\rm{H_2}}$ and
larger PNe have $S_{\rm{Br}\gamma} < S_{\rm{H_2}}$, for the range of
radii shown. The ratios computed from our UWISH2 $1-0$~S(1) surface
brightness and the estimated \brg\ surface brightness are given in
Table~3.

\section{Mid-infrared colours}

\begin{figure*}
\includegraphics[width=17cm,angle=0]{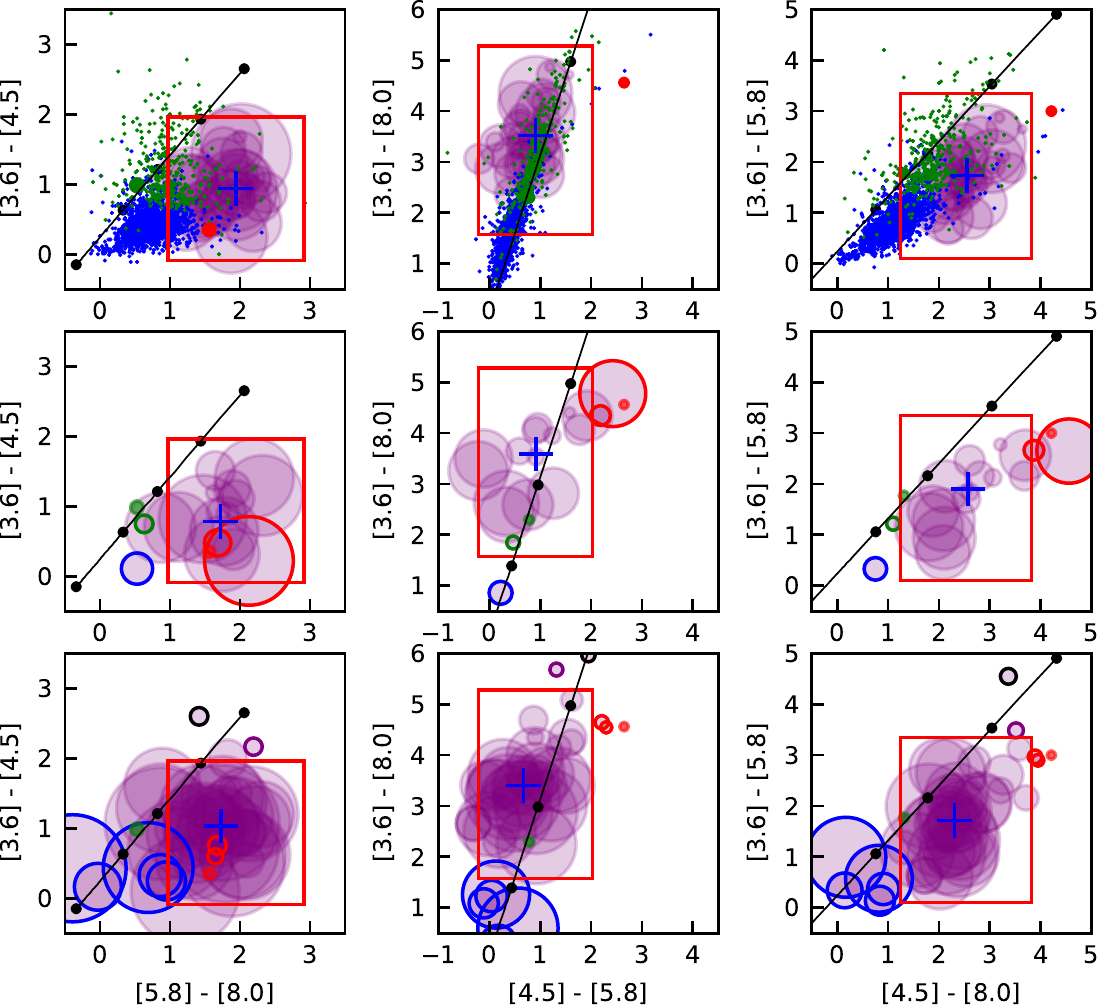}
\caption{{\em IRAC} colour-colour plots for UWISH2 PN candidates shown
  as filled purple circles with circle diameter indicative of the
  photometric errors. {\em top row:} 33 objects flagged as ``True
  PNe'' in the HASH database; {\em middle row:} 18 PN candidates with
  \ha\ detection; {\em bottom row:} 45 PN candidates not detected in
  \ha. Black lines show blackbody colours with temperature varying
  from 300\,K (top right) to 10\,000\,K (bottom left) with points
  marked at 300, 400, 600, 1\,000 and 10\,000\,K. In each plot the
  median colour is shown as a blue cross with $\pm 3$ standard
  deviations in colour shown by the red rectangles. The red and green
  filled circles correspond to PN\,G014.9+00.0 and PN\,G050.4+00.7, a
  \hii\ region and candidate symbiotic system,
  respectively. Additionally in the top row we plot data from Megeath
  et al. (2012) showing the locations of YSOs as small blue and green
  circles (corresponding to disk sources and protostars
  respectively).}
\end{figure*}

Cohen et al. (2011) discuss the {\em Spitzer Space Telescope} IRAC
colours of 136 optically-detected and spectroscopically-confirmed PNe
from the MASH catalogue, as part of a multi-wavelength approach to the
characterization of PNe. They present IRAC colour-colour planes as
diagnostic tools to aid discrimination of PNe from contaminants which
can appear similar at optical wavelengths, such as diffuse and compact
\hii\ regions (see also Zhang et al. 2012). This approach also offers
the possibility to identify PNe on the basis of their mid-infrared
emission, by comparing their IRAC colours with those of
optically-confirmed PNe. For UWISH2 objects this is advantageous given
that the majority of our new PN candidates are not optically visible
and therefore cannot be confirmed using the usual optical diagnostic
lines.

To determine the IRAC colours of our UWISH2 objects we extract
cut-outs from the GLIMPSE
survey\footnote{\url{http://irsa.ipac.caltech.edu/data/SPITZER/GLIMPSE/index_cutouts.html}}
centred on the coordinates of each object, and with a cut-out size
determined by the extent of the \htwo\ emission. Where emission is
seen above the diffuse background level we then measure the flux using
elliptical apertures to encompass all of the emission. The sky
background is estimated by offsetting the aperture to a region of
diffuse background. Any point sources within the apertures are
subtracted from the aperture sum (unless within the object aperture
and associated with the source). We have applied the extended source
correction factors\footnote{The procedure for IRAC extended source
  calibration is given by T. Jarrett at
  \url{http://www.ast.uct.ac.za/~jarrett/irac/calibration/index.html}}
which depend on aperture size; where an aperture radius less than 8
arcsec is used then no correction is applied. Errors on the photometry
are estimated simply from the counts in the object and sky apertures.
Where the emission is faint and/or the diffuse background is high,
then the errors on the IRAC colours will be larger.

We list the IRAC photometry for each object, where available, in
Table\,A1 and give an object ``type''. For known objects which already
have a T, L or P (true, likely or possible PN) designation, from
optical spectroscopy, then this is preserved. For \htwo\ PN candidates
that have no previous observations then by default we list these as
`c' (candidate) unless there is evidence from the IRAC photometry of a
PN nature, in which case we use lower-case t, l, p designations as
follows.  Where photometry is available in all four bands and the six
colours are consistent with PNe (see below) then we list objects as
`t'. Where only three bands are available, but the three resulting
colours are consistent with PNe then we list them as `l'.  If only 2
bands are available but the resulting colour is consistent with PNe
then they are labelled `p'. Where the IRAC colours are more consistent
with a YSO or \hii\ region, then they are labelled as such.

\subsection{True PNe}
There are 76 objects in the UWISH2 sample which are classified as
`true' PNe in the HASH database. The latitude coverage of UWISH2
(generally $|b|\le 1.5$\degr\,over much of the survey, but extending
to higher latitude coverage in places) is larger than that of GLIMPSE
($|b|\le 1.0$\degr) so that only some UWISH2 objects appear in
GLIMPSE.  Of those that do, some do not have usable images in all four
IRAC bands, and in some cases there is no obvious emission above the
background level that can be associated with the object (i.e. lying
within the area defined by the \htwo\ emission). Of the 76 true PNe,
33 have usable IRAC images from which flux in all four bands, and
colour, can be determined.  We plot these objects as purple circles in
the three colour-colour planes in Fig.\,7 (top row), where the PNe are
seen to occupy well-defined regions. The radii of the circles are
equal to the average of the uncertainties in the two colours,
calculated from the photometry. The median colours of the PNe are
marked with blue crosses and are listed and compared to the values of
Cohen et al. (2011) in Table\,4, showing consistent locations; 17/33
of the plotted PNe are present in the MASH and hence common between
the two samples, the rest are in IPHAS.  The likely \hii\ region
PN\,G014.9+00.0 (\#45) and symbiotic system PN\,G050.4+00.7 (\#195)
are also plotted (red and green filled circles respectively) and
occupy regions distinct from the PNe grouping in two of the
three planes.  We define rectangular regions in each colour plane,
shown in red, centred on the median colour and with extent $\pm 3$
standard deviations in each colour. These regions contain all of the
true PNe and are therefore the regions in which PNe are expected to
lie.

In addition to diffuse \hii\ regions, which are known to make up the
majority of detections in UWISH2, the other main contaminant will be
from embedded young stellar objects (YSOs). In drawing up the PN list
we have eliminated objects clearly associated with jets, known
YSOs, or which appear clustered; the IRAC colour
planes offer an additional tool to discriminate between PNe and
YSOs. We plot in Fig.\,7 the locations of YSOs taken from the {\em
  Spitzer} IRAC survey of the Orion A and B molecular clouds by
Megeath et al. (2012). A total of 2\,598 sources are plotted, for
which photometry in all 4 IRAC bands is available, comprising 2\,206
sources identified as pre-main sequence stars with discs and 392
protostars. We take the locations of these objects in the IRAC
colour-colour planes to be typical of the locations of dusty YSOs. The
[5.8]-[8.0] vs. [3.6]-[4.5] plot in particular does a good job of
separating YSOs from PNe, with few YSOs occupying the PN region. The
[4.5]-[5.8] vs. [3.6]-[8.0] and [4.5]-[8.0] vs. [3.6]-[5.8] colour
planes show a good separation between PNe and the likely location of
diffuse \hii\ regions (red circle), with slightly more contamination
of the PN region by YSOs, especially in the former case where the
track of cool blackbody colours runs through the PN region.

\subsection{Other optically-visible PN candidates}
The IRAC colours of 18 PN candidates with detected \ha\ emission, but
which are not spectroscopically confirmed as true PNe, are shown in
the middle row of Fig.\,7. These include objects with `L' or
`P' designations as well as objects newly detected in UWISH2. 

One object, \#234 (PN\,G059.7-00.8), lies outside the PN box in all
three planes to the bottom left, in the region occupied by YSOs
(blue-edged circle). The object has a bipolar (Bs), spider-like
morphology in \htwo\ (Fig.\, B1) with a central peak.  In IPHAS
\ha\ imaging it appears as a compact bright central region with a
fainter halo. It is identified spectroscopically as a likely PN by
Sabin et al. (2014).  We have obtained {\em K}-band spectroscopy with
{\em LIRIS} on the WHT which shows \htwo\ line ratios characteristic
of thermal excitation in the nebula and evidence for the $v=2-0$ and
$3-1$ CO bandheads in absorption at the central position (Jones et
al. 2018, submitted).  CO bandhead absorption is seen in many YSOs,
where it can arise in the photosphere (e.g. Casali \& Eiroa 1996) or
in the disc, especially in FU Ori type systems (Reipurth \& Aspin
1997). CO absorption is also seen in the spectra of cool evolved
objects, such as pre-PNe with spectral types G and later, but tends to
be absent in hotter post-AGB objects (e.g.  Hrivnak, Kwok \& Geballe
1994). It is possible that PN\,G059.7-00.8 is a YSO, but given its
current PN status it seems more likely to be a PN in chance alignment
with a field star.

Object \#283 lies outside the PN box in two colour planes and close to
the blackbody track; it also lies close to the position occupied by
PN\,G050.4+00.7, in all three colour planes (green-edged circle),
which is listed as a possible symbiotic star in the HASH database,
although Sabin et al. (2014) list it as a likely PN.  The object has
irregular (I) morphology in \htwo\ with a compact central source
visible in the SHS \ha\ image (Fig.\, B4). Its location in Fig.\,7
may, therefore, indicate a symbiotic (or possibly YSO) nature rather
than a PN, however we note that it is often difficult to differentiate
between the two classes (Frew \& Parker 2010). A particular case is
Mz\,3, which has been classed as PN and symbiotic and has a
mid-infrared-to-radio flux ratio characteristic of \hii\ regions
(Cohen et al. 2011).

Two objects (red-edged circles) have colours similar to those of
diffuse \hii\ regions.  These are \#038 and \#171 which have bipolar
(Bp) and round (Rs) \htwo\ morphologies (Figs\,B5 and B7
respectively). The latter object is an IRAS source - IRAS 19117+0903.

\subsection{Optically invisible PN candidates}  

There are 45 UWISH2 PN candidates with no \ha\ detection and for 
which IRAC photometry in all 4 bands is obtained. These are shown 
in the bottom row of Fig\,7. We note that the majority of objects
have IRAC colours typical of PNe and that the median colours are
well-centred in the box defined by the ``True'' PNe, so that our
initial selection procedure (based on morphology and avoidance of
star-forming clouds) was mostly successful.

A group of 5 objects lies to the bottom left of the colour planes,
outside the PN box and in the region occupied by YSOs. These are again
shown as blue-edged circles and are \#014, \#111, \#124, \#012 and
\#032. Three objects (\#014, \#111 and \#124; Fig.~C4) appear
asymmetric in \htwo\ emission whereas two (\#012 and \#032; Fig.~C1)
are possibly bipolar. {\em K}-band spectroscopy of \#012 shows
\htwo\ line ratios typical of thermal excitation in the nebula, with
no evidence of \brg\ emission (Jones et al. 2018, submitted)
and so is a likely YSO. The remaining 4 objects could
also be YSOs, although the IRAC images contain point sources so it is
possible that these are also PNe superimposed with field stars.

Two objects have colours similar to those of diffuse \hii\ regions
(red-edged circles). These are \#029 and \#048 (Figs.~C2 and C1
respectively). Both are bright in the IRAC bands and we consider them
as likely \hii\ regions.

Two objects lie outside the PN box in the upper part of the colour
planes.  The object with the more extreme colours is \#066, shown with
a black-edged circle. In \htwo\ imaging it appears as a compact
bipolar consisting of two small blobs of emission (Fig.~C1) with
extent $2.8\times 1.2$ arcsec. {\em K}-band spectroscopy shows no
\brg\ emission (Jones et al. 2018, submitted).  Given the small
dimensions and lack of evidence of ionization, this is likely to be a
pre-PN. Object \#114 is shown with a purple-edged circle and also lies
outside the PN box. This object appears as an X-shaped bipolar in
\htwo\ imaging, with a bright central star. It is a radio and
millimetre source and has been listed as a PN (Urquhart et
al. 2009). In a survey of the inner GP for Wolf-Rayet stars, Kanarek
et al. (2015) also class this object as a possible PN (object
1527-318B); their near-infrared spectrum shows strong emission from
\hei\ (2.058\,$\umu$m) and \brg\, which can be characteristic of young
PNe (e.g. Gledhill \& Forde 2015).  These lines are also found in
massive YSO spectra and \hii\ regions, but in the case of MYSOs the
continuum rises steeply into the red (Cooper et al. 2013), whereas the
continuum of \#114 appears flat. The IRAC images show a compact source
suggesting that the object is not a diffuse \hii\ region either.
Although \#114 lies outside the PN box in our colour-colour plots, we
retain it as a candidate PN.

\renewcommand{\arraystretch}{0.8}
\begin{table}
\caption{Median IRAC colours and standard error on mean (sem) for 33 UWISH2 PNe classed 
as ``True'' PNe, and the set of ``All PNe'' from the MASH sample of Cohen et al. (2011).}
\begin{tabular}{lccccc} \\
\hline
Colour       & \multicolumn{3}{c}{This work} & \multicolumn{2}{c}{MASH Sample}  \\
             & median & std. dev. & sem & median & sem \\
\hline
$[3.6]-[4.5]$ & 0.94 & 0.34 & 0.06 & 0.81 & 0.08 \\
$[3.6]-[5.8]$ & 1.73 & 0.54 & 0.10 & 1.73 & 0.10 \\ 
$[3.6]-[8.0]$ & 3.43 & 0.62 & 0.11 & 3.70 & 0.11 \\ 
$[4.5]-[5.8]$ & 0.91 & 0.37 & 0.07 & 0.86 & 0.10 \\
$[4.5]-[8.0]$ & 2.53 & 0.43 & 0.07 & 2.56 & 0.11 \\
$[5.8]-[8.0]$ & 1.94 & 0.32 & 0.06 & 1.86 & 0.07 \\
\hline
\end{tabular}
\end{table}  
\renewcommand{\arraystretch}{1.0}

\section{PNe Towards Open Clusters}

A possible, although rarely available, approach to measuring the
distance to a PN, along with the age of the PN progenitor, is to
identify PNe in open clusters. Several studies of this type have been
made but the number of confirmed cluster members is small. A summary
is given by FPB16, who use 9 PN-cluster pairings as calibrators for
their $S_{\rm H\alpha}-r$ relation, including 4 PNe in globular
clusters.  We have hence cross-checked our list of PN candidates to
identify objects that potentially could belong to open clusters.

We utilised the Milky Way Star Cluster (MWSC) catalogue by Kharchenko
et al. (2013) for this purpose. This catalogue features a list of 3006
confirmed clusters (open, globular and associations) from a number of
sources and provides a uniform and homogeneous set of cluster
parameters such as positions, radii, distances and ages. We find a
total of 18 PN candidates which are within the angular radius of one
of the star clusters. The full list of objects and the associated
clusters and their properties are shown in Table~5.

Some of these associations could be chance projections.  In order to
evaluate what fraction of sources could be real associations we run
some basic simulations. This was only done for the core region of
UWISH2, i.e. for the objects at Galactic longitudes between 0\dg\ and
65\dg and Galactic latitudes between $\pm$1.7\dg. This region has a
homogeneous coverage in UWISH2, in both coordinates. There are 15 PNe
which overlap with clusters in this area.  The remaining three
associations of PNe and clusters occur in the additional UWISH2 fields
taken in Cygnus and Auriga/Cassiopeia where no homogeneous coverage
has been achieved.

\begin{figure}
\includegraphics[clip,trim=-2cm 0.0cm 6cm 11cm, width=12cm,angle=0]{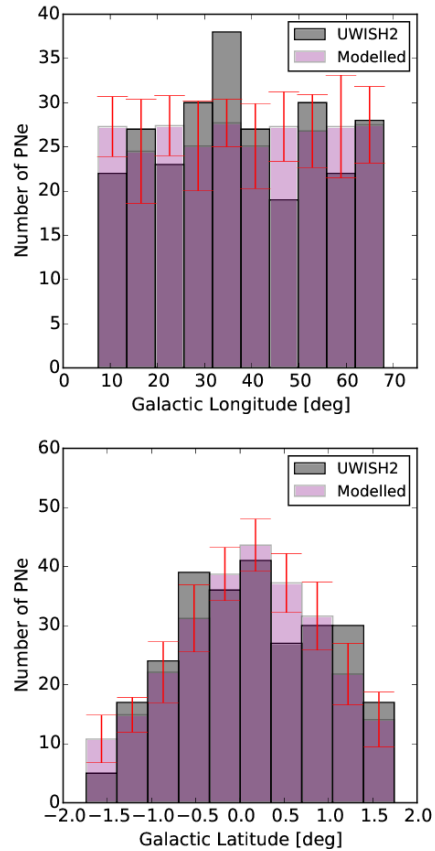}
\caption{Observed and modelled planetary nebulae Galactic distribution
histograms from UWISH2. The purple bins represent the average modelled
distribution from 1000 simulations, with $\pm 1\sigma$ error bars. The
grey bins are the observed PNe from the survey ($N_{\rm PNe}=266$).}
\end{figure}

In order to estimate how many associations of PNe and clusters one
would expect, we randomly distribute the same number of PNe in the
survey area and identify how many of them are projected onto the star
clusters. We repeat this simulation 1000 times to estimate a reliable
mean and uncertainty for our estimates. We also ensure that the random
positions in the simulation follow the same distribution as the real
PNe. In other words the Galactic longitudes are sampled from a
homogeneous distribution, while the Galactic latitudes are sampled
from a Gaussian distribution with a width of 0.83\dg\ and a centre
position of -0.03\dg. In Fig.\,8 we show the observed distributions in
$l$ and $b$, as well as the average and scatter of the simulated
distributions. We do not randomise the positions of the clusters
during the simulation, but keep them fixed in their real
positions. The reason for this that the MWSC list has to a large
extent been made by combining cluster lists from the
literature. Hence, the cluster list will suffer from a non-homogeneous
coverage. Thus, randomising the position may cause erroneous results.

We find that on average in our 1000 simulations there are
13.2\,$\pm$\,3.6 PN projected onto one of the clusters. Hence, within
the uncertainties the 15 PNe that overlap with clusters are to be
expected randomly. Thus, the number of PNe that are expected to be
associated with a cluster seems to be very small. We attempted to run
a larger number of simulations to reduce the uncertainties, but find
that the scatter is not reduced once the number of runs is above
1000. We also slightly changed the latitude distribution of the PNe in
the simulation but obtain the same results for the number of
associations within the error bars.

One would expect that PNe in clusters occur mostly once the age of the
cluster exceeds $\sim50$\,Myr, which corresponds approximately to the
time for a 7~M$_{\odot}$ star to evolve to the post-AGB stage
(e.g. Choi et al. 2016).  Out of 15 possible associations of PNe with
clusters, 5 are for clusters with ages above 100\,Myr. These five (in
the clusters Ruprecht\,137, ASCC\,94, Kharchenko\,2, Kronberger\,2,
NGC\,6649) may then be the most likely objects to be really associated
with clusters.

\begin{table*}
\caption{UWISH2 objects projected onto open clusters showing object
  number (Table\,A1) and Galactic coordinates. The cluster id, name,
  angular radius, estimated distance, colour excess and age (from
  Kharchenko et al. 2013) are given. $\Delta r_{2}$ is the separation
  of the PN candidate from the cluster centre in units of $r_{2}$. The
  final column gives the object type, as listed in Table\,A1. The
  first 5 objects are projected onto clusters with $\log{(\rm
    age})>7.7$\,Myr.}
\begin{tabular}{lHcccccccccc}
\hline
Object \#&PN\,G     &$l$ (deg)&$b$ (deg)& MWSC& Cluster& $r_{2}$\,(deg) & $D$ (pc)& $E(B-V)$& $\log(t)$\,(Myr)&$\Delta r_{2}$& Type \\
\hline
012&004.7-00.8 & 4.76799 & -0.85257 &
2772 & Ruprecht 137 &
0.16 &
1405 & 1.124 & 8.80 &
0.34 & YSO \\
290& & 15.46686 &1.10391 &
2858 & ASCC 94 &
0.27&
731 &0.25 &8.84 &
0.76 & c \\
071&021.7-00.6 & 21.74338 &-0.67287 &
2949 &NGC 6649 &
0.18&
1564 & 1.332 &8.1 &
0.82 & T \\
056&016.9-00.0a &16.92321 &-0.00616 &
2898 &Kronberger 2 &
0.14&
3197& 1.791 &8.2 &
0.77 & c \\
055&016.6-00.2 &16.60078 &-0.27565 &
2900 &Kharchenko 2 &
0.19&
3412 &1.041 &8.43 &
0.48 & t \\
\hline
018&005.9-01.3 & 5.90685 & -1.37325 &
2798 & NGC 6530 &
0.2&
1365 & 0.541 & 6.67 &
0.89 & c \\
286&006.3-00.6 & 6.33697 & -0.6148 &
2781 & Bochum 14 &
0.167&
538 & 1.853 & 7.1 &
0.67 & c \\
021&008.3-01.1 &8.33574 &-1.10291 &
2814 &ASCC 93 &
0.165&
1830 & 0.583 & 6.1 &
0.38 & T \\
027&010.2+00.3 & 10.21147 &0.34469 &
2809 &FSR 0039 &
0.14&
2940 &0.802 &6.0 &
0.30 & T \\
046&015.1-00.4 &15.13012 &-0.44046 &
2896 &NGC 6618 &
0.29&
1308 &1.607 & 6.0 &
0.88 & t \\
060&018.1+01.5 &18.14941 &1.53214 &
2878 &NGC 6604 &
0.265&
1895 & 0.968 &6.89 &
0.74 & T \\
084&024.8+00.4 &24.89596 &0.45853 &
2957 &Dolidze 31 &
0.115&
5163 &1.353 &6.0 &
0.77 & t \\
090&026.4-00.8 &26.44767 &-0.8084 &
2977 &NGC 6683 &
0.11&
1441 &0.333 &6.75 &
0.79 & c \\
117&031.8+00.8 &31.16908 &0.81029 &
2986 &Berkeley 79 &
0.11&
2434 &1.145 &6.8 &
0.99 & c \\
303& &60.02622 &-0.28202 &
3173 &Roslund 2 &
0.18&
1668 &0.899 &6.78 &
0.77 & c \\
258&075.9+00.2 &75.90338 &0.29517 &
3325 &Berkeley 87 &
0.18&
1239 &1.353 &7.1 &
0.95 & c \\
274&143.5-02.8 &143.5014 &-2.81706 &
274 &Melotte 20 &
6.1&
175 &0.09 &7.7 &
0.89 & poss \hii \\
278&149.4-02.1 &149.43257 &-2.19327 &
274 &Melotte 20 &
6.1&
175 &0.09 &7.7 &
0.78 & c \\
\hline
\end{tabular}
\end{table*}

\#012: IRAC colours suggest that this object is likely to be pre-main sequence
(Sec.\,5.3). Distance and reddening
are unknown although a median $E(B-V)=2.6$ along this line of sight 
(Sec.\,3.4.1) implies that it may be background to the cluster. 

\#290: A small (5 x 3 arcsec) bipolar, possibly associated with 
ASCC\,94 although the cluster distance would imply an object extent of less than 
$0.01$~pc, more typical of a pre-PN.

\#071: A confirmed PN (PN\,G021.7-00.6; M\,3-55) with $S_{H\alpha}-r$ distance of 
$5.86\pm2.46$\,kpc and $E(B-V)=1.61$ (FPB16) suggesting it is background to the cluster,
even assuming an error on the cluster distance as large as 50 per cent. 

\#056: A faint 8 arcsec-diameter ring of \htwo\ emission, possibly
associated with Kronberger\,2, although the median reddening along this line of sight is
$E(B-V)=3.4$ so that most stars are background to the cluster.

\#055: This object appears as a small (7 x 7 arcsec) bipolar in UWISH2
images and is projected within the cluster radius of Kharchenko
2. However it lies close ($\sim 2$ arcmin) to the western edge of the
\hii\ region Sh2-48 and the IR bubble N18 which are associated with
ongoing star formation (Ortega et al. 2013; Deharveng et al. 2010). It
is coincident with a compact radio source GPSR5\,16.600-0.276 (Becker
et al. 1994) and was identified as a possible compact \hii\ region by
Deharveng et al. (2010).  The radio emission is resolved in the 5\,GHz
CORNISH survey (Purcell et al. 2013) with an integrated flux of
$22.2\pm4.3$\,mJy, angular scale 4.2 arcsec, and a morphology similar
to that seen in \htwo\ emission. The mid-infrared colours from IRAC
photometry (Table\,5; Sec.\,4) are more typical of PNe than
ultra-compact \hii\ regions so we consider this object a likely
PN. The extinction map of Rowles \& Froebrich (2009) give $E(B-V)=1.1$
for this object, which is similar to that reported for the cluster. If
this is really a PN then this is the most likely candidate for cluster
membership in our sample. 

Objects projected onto one of the young clusters may be candidate YSO
outflows, rather than PNe. Three of the objects are confirmed PNe (T)
and two have IRAC colours consistent with PNe (t), so these must be
chance alignments. There are a further eight objects; seven candidate
PNe (c), and one (\#274) which is likely to be associated with the
\hii\ region BFS\,30.

\section{Discussion and summary}
The 284 PN candidates presented in F15 are supplemented with a further
23 (mostly very faint) \htwo\ nebulae (Table\,A2) to give a final
list of 307 objects. Cut-out images showing the \htwo\ emission are
given in Appendix B and C and all objects are listed in Table\,A1 along
with their \htwo\ morphology and dimensions.

Where available we have obtained mid-infrared photometry in the four {\em
  Spitzer} IRAC bands to place objects onto diagnostic colour-colour
diagrams. This has allowed us to identify 11 objects which are
probably not PNe: 5 YSOs, 4 \hii\ regions, 1 symbiotic system and 1
pre-PN (Sec.\,4). Adding in 5 objects which have previously been
classified as not PNe, there are 16 non-PNe in the sample, leaving a
total of 291 PN candidates, 183 of which are new detections. For PN
candidates with IRAC colours, we find that they are well-clustered
within the regions in the colour-colour planes defined by the `true'
PNe (Fig.\, 7), showing that our initial selection of objects, based
mainly on appearance in \htwo\ imaging, was largely successful. We
have used the classifiers t, l and p, to indicate the likeliness of a
PN nature, on the basis of the IRAC colours (Sec.\,5 and Table\,A1).

By searching for \ha\ emission at the location of the UWISH2 sources,
in either the SHS or IPHAS surveys, we find that 95 of the 108
previously known PN candidates show \ha\ emission (Figs.\,B1-B5). This
is not surprising given that the main discovery technique for PNe has
been by identification in \ha\ surveys. In addition we find 28 objects
that appear to be associated with \ha\ emission but which have not
been previously noted (Figs.\,B6-B8). The majority of the new PN
candidates identified by their \htwo\ emission, 155 out of 183, have
no detection in \ha\ emission, raising the possibility of a
significant population of optically-obscured PNe.

We find in Sec.\,3.4 that the spatial distribution of PN candidates
with and without detectable \ha\ emission is noticeably different,
with there being fewer \ha-detected PNe in the inner GP ($l<20$\degr)
and at lower latitudes ($-1\degr < b < 1\degr$). To assess the effects
of extinction on the ability to detect PNe at optical wavelengths, we
have used the extinction maps of Rowles \& Froebrich (2009) to obtain
estimates of $A_{\rm v}$ at the position of each object. We find that,
in general, the objects with no detectable \ha\ emission are located
in regions with higher extinction than those where \ha\ emission is
detected, with median values of $A_{\rm v}=5.4$ and 7.2 mag
respectively for the two groups. We also find that objects without
detectable \ha\ emission typically have smaller angular extent on the
sky than those with detectable \ha\ emission (median surface area of
70 and 160 arcsec$^2$, respectively), suggesting that the former group
are located at greater distance and therefore suffer more
extinction. The flux distributions of the two groups are also distinct
(Fig.\,4), with median fluxes of 9.1 and $3.2\times
10^{-17}$\,W\,m$^{-2}$ for objects with and without
\ha\ detection. The combined trends of lower flux and smaller angular
extent, with greater extinction, suggest that UWISH2 is probing longer
sightlines in the GP to detect a population of new and faint objects
that are likely to be optically-obscured PN candidates.

We have compared the \ha\ and \htwo\ surface brightnesses for 23 PNe
for which physical radii are available from the $S_{\rm H\alpha}-r$
relationship. Due to the flat trend of $S_{\rm H_2}$ with radius, the
\htwo\ and \ha\ surface brightnesses become comparable for PNe with
radii greater than 0.5\,pc (Fig.\,6).  The maximum radius of a visible
PN is estimated as 0.9\,pc, beyond which it falls below
\ha\ detectability at age 21\,000$\pm$5\,000 years (Jacob et al.
2013). It seems likely that older PNe will be more easily detected via
their \htwo\ emission, especially in the GP where extinction at
optical wavelengths is high. For PNe which show no detectable
\htwo\ emission then other diagnostics may be useful. The MIPSGAL
24\,$\umu$m survey is thought to contain many potential PNe (Mizuno et
al. 2010) and in some cases the emission is dominated by the [O{\sc
    iv}] line at $25.9 \umu$m, suggesting that they are
high-excitation PNe (Nowak et al., 2014; Flagey et al. 2011).

To assess the PN detection efficiency of UWISH2 we search for known
PNe within the area $10\degr < l < 66\degr$ and $|b|<1.5\degr$ using
the HASH database, giving a total of 287 objects. Of these, 85 are
detected in UWISH2 (30 per cent). However, if we take only objects
with reliable spectroscopic confirmation (`true' PNe) then 61 out of
114 objects (54 per cent) are found in UWISH2 (Sec.\,3.3). The
probability of detecting \htwo\ emission from a PN would seem to lie
between 30 and 54 per cent. Within the same area we detect 131 PN
candidates in UWISH2 which do not appear in \ha\ surveys. Assuming
that these objects are all optically-obscured PNe then, with a
detection efficiency of 30 or 54 per cent, there would be in total 437
or 243 PNe hidden from \ha\ surveys. Including infrared detections can
therefore be expected to significantly increase the number of PN
candidates in a given area of sky, especially close to the GP where
extinction effects are most severe. In the sample area ($10\degr < l <
66\degr$, $|b|<1.5\degr$) the number increases from 287
optically-detected PNe to at least 418 (287+131) objects, but up to
724 (287+437). Including near-infrared detections increases PN
candidate numbers by a factor between 1.5 and 2.5 within the UWISH2
survey area.

\section*{acknowledgements}
This paper makes use of data obtained as part of the INT Photometric
\ha\ Survey of the Northern GP (IPHAS, www.iphas.org) carried out at
the Isaac Newton Telescope (INT). The INT is operated on the island of
La Palma by the Isaac Newton Group in the Spanish Observatorio del
Roque de los Muchachos of the Instituto de Astrofisica de
Canarias. All IPHAS data are processed by the Cambridge Astronomical
Survey Unit, at the Institute of Astronomy in Cambridge. This research
has made use of: the NASA/IPAC Infrared Science Archive, which is
operated by the Jet Propulsion Laboratory, California Institute of
Technology, under contract with the National Aeronautics and Space
Administration; the SIMBAD database, operated at CDS, Stasbourg,
France; the HASH PN database at `hashpn.space'. AMJ acknowledges
receipt of a STFC research studentship.  We thank the referee for
their helpful comments.

\label{lastpage}

\onecolumn

\setcounter{table}{0}
\renewcommand{\thetable}{A\arabic{table}}
\begin{table*}
\flushleft{\bf APPENDIX A: PN CANDIDATE LISTS}
\setlength{\tabcolsep}{5pt}
\renewcommand{\arraystretch}{0.8}
\caption{Morphology, extent and IRAC photometry for UWISH2 PN
  candidates. Columns are (1) candidate number; (2) UWISH2 ID; (3)
  morphology and (4) extent of \htwo\ emission (major x minor axis
  dimensions in arcsec, or where only one dimension is given it is the
  major axis); (5) morphology and (6) extent of any \ha\ emission; (7)
  to (10) IRAC magnitudes and errors; (11) object type: where an
  existing classification (TLP) exists then this is preserved. For new
  objects: t, l, p, c indicates ``true'', ``likely'', ``possible'' or
  ``candidate'' PN status based on the IRAC colours where photometry
  in 4, 3, 2 or $\le 1$ bands is available, respectively.}
\begin{tabular}{llccccccccr}
\hline
No.&UWISH2 ID &  \multicolumn{2}{c}{\htwo} & \multicolumn{2}{c}{\ha} & \multicolumn{4}{c}{IRAC mag.}  & Type \\
   &       &    Morph. & Size           & Morph. & Size           & [3.6] & [4.5] & [5.8] & [8.0]  &  \\
\hline
001&000.81878-0.04944 & Ear   & 5x4  &  &  &  12.42 $\pm$ 0.10 & 11.34 $\pm$ 0.07 & 9.80 $\pm$ 0.06 & 8.08 $\pm$ 0.04&  	t \\
002&001.22588+0.56414 & Ear & 3.0x2.6 &  & & & & & &	c \\
003&001.42213-0.61357 & Bamr & 22x12 &  & &  &  &  &  &	c \\
004&001.65056+0.18803 & Barm & 26x13 & E & 12x6 & 12.34 $\pm$ 0.10 & 11.83 $\pm$ 0.09 & 10.82$\pm$0.11& 9.59$\pm$0.11 &	T \\
005&001.72196-0.82262 & Eamr & 20x12 &  & & & & & & 	c \\
006&001.73118+0.44232 & Brs & 8x5 &  & &  &  &  &  &	c \\
007&002.03824-0.34363 & Ear & 34x18 &  &  & & & & &	c \\
008&002.25319+0.55724 & Ears & 30x20 & Eprs & 38x12&  &  &  &  &	T \\
009&003.65197-0.25068 & Ear & 7x5 &  &  & & & & &	c \\
010&003.79367-0.81428 & Ear & 4x3 &  &  & & & & &	c \\
011&004.44847-0.21905 & Ims & 120x90 &  &  & & & & &	c \\
012&004.76799-0.85257 & Bams & 25x12 &  & &12.71 $\pm$ 0.11 & 12.27 $\pm$ 0.11 & 12.13 $\pm$ 0.22 & 11.45 $\pm$ 0.44& 	YSO \\
013&004.88887-0.58981 & Is/B & 34x12 & Ias/B & 88x62&  &  &  &  &	T \\
014&004.99841-0.72107 & A & 9x4 &  & &9.88 $\pm$ 0.02 & 9.71 $\pm$ 0.03 & 9.53 $\pm$ 0.05 & 9.56 $\pm$ 0.15 &	YSO? \\
015&005.14078+0.58616 & Rar & 7 &  &  & & & & &	c \\
016&005.42811-0.16852 & Bms & 40x9 & B & 10x3 & & & & 8.03 $\pm$ 0.04 &	c \\
017&005.56222-0.15023 & Rar & 7 &  &  & & & & &	c \\
018&005.90685-1.37325 & Ear & 8x5 &  &  & & & & &	c \\
019&006.51856-0.69279 & Ear & 10x6 &  &  & & & & &	c \\
020&006.73073-1.19177 & Bas & 9x6 &  & &12.66 $\pm$ 0.15 & 12.03 $\pm$ 0.12 & 12.05 $\pm$ 0.35 & 10.01 $\pm$ 0.17&	t \\
021&008.33574-1.10291 & Is & 35 & Emps & 30x20&  &  &  &  &	T \\
022&008.36142-0.62384 & Bmps & 8x6 &  & &  &  &  &  &	c \\
023&008.94169+0.25318 & Ears & 20x10 & Er & 27x11 & & & &11.94 $\pm$ 1.50 &	L \\
024&009.76150-0.95756 & Bs & 18x6 &  & &11.97 $\pm$ 0.07 & 9.98 $\pm$ 0.03 & 9.11 $\pm$ 0.03 & 7.29 $\pm$ 0.02&	t \\
025&009.80708-1.14613 & Bmps & 36x14 & Bms & 32x12 & 11.82 $\pm$ 0.08 & 10.32 $\pm$ 0.04 & 9.27 $\pm$ 0.05 & 7.51 $\pm$ 0.03 &	T \\
026&010.10373+0.73752 & Bmps & 140x36 & Bmps & 11x10 & 7.66 $\pm$ 0.01 & 5.99 $\pm$ 0.00 & 4.80 $\pm$ 0.00 & 2.80 $\pm$ 0.00 &	T \\
027&010.21147+0.34469 & Brs & 56x20 & Bams & 61&  &  &  &  &	T \\
028&010.26120-0.79452 & Bmr & 12x7 &  & & 2.02 $\pm$ 0.06 & 11.77 $\pm$ 0.10 & 10.33 $\pm$ 0.15 & 8.91 $\pm$ 0.11& 	t \\
029&010.39239+0.53966 & Emr & 24x13 &  & &8.08 $\pm$ 0.01 & 7.49 $\pm$ 0.01 & 5.19 $\pm$ 0.00 & 3.54 $\pm$ 0.00&	\hii \\
030&010.94194-0.40277 & Bas & 15x13 &  & &12.09 $\pm$ 0.12 & 10.99 $\pm$ 0.07 & 10.91 $\pm$ 0.22 & 8.69 $\pm$ 0.10&	t \\
031&011.00185+1.44395 & Bas & 26x16 & E & 19x17 &10.74 $\pm$ 0.05 & 9.92 $\pm$ 0.03 & 8.79 $\pm$ 0.03 & 6.88 $\pm$ 0.02&	T \\
032&011.32982+0.54981 & B & 11x5 &  & &9.66 $\pm$ 0.03 & 9.34 $\pm$ 0.03 & 9.30 $\pm$ 0.06 & 8.44 $\pm$ 0.08&	YSO? \\
033&011.45829+1.07349 & Ar & 6x3.5 &  & &  &  &  &  &	P \\
034&011.52915+1.00385 & R & 9 & E & 6.5x6.0 & & & & &	P \\
035&011.86338+0.30190 & E & 10x8 &  &  & & & & &	c \\
036&012.11515+0.07516 & Bbs & 22x10 &  & &10.11 $\pm$ 0.02 & 9.07 $\pm$ 0.02 & 8.79 $\pm$ 0.03 & 6.93 $\pm$ 0.02&	L \\
037&012.20907+0.43081 & I & 11x5 &  &  & & & & &	c \\
038&012.21971-0.33477 & Bp & 16x5 & A & 20x10&13.89 $\pm$ 0.27 & 13.67 $\pm$ 0.31 & 11.25 $\pm$ 0.13 & 9.12 $\pm$ 0.12& \hii \\
039&012.71728+0.37202 & Bp & 14x10 &  & &12.59 $\pm$ 0.21 & 11.19 $\pm$ 0.11 & 10.82 $\pm$ 0.22 & 9.01 $\pm$ 0.11& 	t \\
040&012.80348+0.00510 & Er & 16x14 &  &  & & & & &	c \\
041&013.61090+1.01274 & Ear/Be & 13x8 & R & 13x13 & &11.69 $\pm$ 0.08  & &10.08 $\pm$ 0.19 &	p \\
042&014.58523+0.46161 & B & 9x4 & Ea & 6x5&9.76 $\pm$ 0.02 & 8.65 $\pm$ 0.01 & 8.53 $\pm$ 0.02 & 6.36 $\pm$ 0.01&	L \\
043&014.64501+0.08920 & Ba & 14x8 & E & 20x10&10.97 $\pm$ 0.05 & 10.59 $\pm$ 0.06 & 8.81 $\pm$ 0.04 & 6.91 $\pm$ 0.02&	t \\
044&014.65833+1.01220 & Ears & 22x20 & Eas & 30x22&  &  &  &  &	T \\
045&014.92112+0.06989 & Ers & 30x16 & R &8 & 7.06 $\pm$ 0.01 & 6.71 $\pm$ 0.01 & 4.07 $\pm$ 0.00 & 2.51 $\pm$ 0.00&	\hii \\
046&015.13012-0.44046 & Bas & 40x20 &  & &11.19 $\pm$ 0.07 & 10.02 $\pm$ 0.05 & 9.33 $\pm$ 0.07 & 7.66 $\pm$ 0.05&	t \\
047&015.53753-0.01923 & Er & 35x7 & Emr & 55x11 & & & & &	L \\
048&015.54859-1.00657 & Bps & 12x3 &  & &8.89 $\pm$ 0.01 & 8.13 $\pm$ 0.01 & 5.92 $\pm$ 0.01 & 4.25 $\pm$ 0.00& 	\hii \\
049&015.67993-1.36320 & Ear & 12x10 &  &  & & & & &	c \\
050&016.02790-1.00525 & Rar & 7  &  &  & & & & &	c \\
051&016.11984-0.98789 & B & 30x15 &  & &  &  &  &  &	c \\
052&016.17480+1.37914 & Bp & 24x10 &  & &  &  &  &  &	c \\
053&016.41571-0.93047 & Bpr & 54x32 & Bps & 30x18&  &  &  &  &	T \\
054&016.48834-1.36082 & Is & 45x40 &  &  & & & & &	c \\
055&016.60078-0.27565 & Bs & 7x7 &  & &12.11 $\pm$ 0.12 & 10.91 $\pm$ 0.06 & 11.00 $\pm$ 0.18 & 8.70 $\pm$ 0.08&	t \\
056&016.92321-0.00616 & Rar & 8 &  &  & & & & &	c \\
057&017.22288+0.12645 & Bs & 36 & B & 20x6 & & & & &	L \\
058&017.58861+1.09048 & Bas & ~9x4 & S & &13.23 $\pm$ 0.17 & 12.08 $\pm$ 0.09 & 12.29 $\pm$ 0.26 & 9.97 $\pm$ 0.15&	t \\
059&017.61528-1.17013 & As/Bas? & 240 & E/B & & 7.23 $\pm$ 0.01 & 6.58 $\pm$ 0.01 & 5.52 $\pm$ 0.01 & 4.12 $\pm$ 0.01 &	T \\
\end{tabular}
\end{table*}
\renewcommand{\arraystretch}{1.0}

\renewcommand{\arraystretch}{0.8}
\setcounter{table}{0}
\begin{table*}
\caption{continued.}
\begin{tabular}{llccccccccr}
\hline
No. & UWISH2 ID &  \multicolumn{2}{c}{\htwo} & \multicolumn{2}{c}{\ha} & \multicolumn{4}{c}{IRAC mag.}  & Type \\
    &      &    Morph. & Size           & Morph. & Size           & [3.6] & [4.5] & [5.8] & [8.0]  &  \\
\hline
060&018.14941+1.53214 & Bps & 16x10 & Bams & 16x11&  &  &  &  &	T \\
061&018.41760-0.10793 & B & 16x12 &  & &  &  &  &  &	c \\
062&018.83207+0.48278 & R & 3.5 &  & &14.50 $\pm$ 0.32 & 13.44 $\pm$ 0.22 & 12.51 $\pm$ 0.32 & 11.64 $\pm$ 0.44&	t \\
063&020.46958+0.67836 & Rar & 40 & B/R & 8x7&10.22 $\pm$ 0.02 & 9.15 $\pm$ 0.02 & 8.55 $\pm$ 0.02 & 6.28 $\pm$ 0.01&	T \\
064&020.70907-0.17267 & Bs & 14x9 & A & 15x9 &12.28 $\pm$ 0.12 &11.80 $\pm$ 0.14 & & &	p \\
065&020.80590-0.57267 & Bprs & 36x16 &  & &  &  &  &  &	c \\
066&020.85450+0.48588 & B & 2.8x1.2 &  & &10.08 $\pm$ 0.02 & 7.49 $\pm$ 0.01 & 5.54 $\pm$ 0.00 & 4.12 $\pm$ 0.00& 	pre-PN \\
067&020.97795+0.92363 & Bms & 35x5 & E & 10x8&11.47 $\pm$ 0.04 & 10.36 $\pm$ 0.03 & 9.76 $\pm$ 0.04 & 7.82 $\pm$ 0.02&	L \\
068&020.98141+0.85244 & Is & 35x16 & Aa & 47x11 & & & &8.95 $\pm$ 0.14 & 	L \\
069&021.29383+0.98091 & Is/Bs & >60 & Bs & 54x42&  &  &  &  &	T \\
070&021.30767-0.25089 & Is & 25x25 &  &  & & & & &	c \\
071&021.74338-0.67287 & Brs & 22x7 & B & 12x9&12.17 $\pm$ 0.11 & 11.23 $\pm$ 0.09 & 10.44 $\pm$ 0.15 & 9.22 $\pm$ 0.15&	T \\
072&021.81951-0.47837 & Bprs & 52x12 & Bp & 24x12&10.61 $\pm$ 0.04 & 9.85 $\pm$ 0.04 & 8.73 $\pm$ 0.06 & 6.44 $\pm$ 0.03&	T \\
073&022.44734-0.44228 & Ears & 32x26 &  &  & & & & &	c \\
074&022.57000+1.05505 & Brs & 22x15 & Bs & 23x17&10.55 $\pm$ 0.03 & 9.79 $\pm$ 0.03 & 9.20 $\pm$ 0.04 & 7.24 $\pm$ 0.02&	T \\
075&022.99501-0.56968 & I & 10x6 &  & &  &  &  &  &	c \\
076&022.99982+0.10714 & Bps & 40x22 &  &  & & & & &	c \\
077&023.44011+0.74528 & Bs & 10x5 & Eas & 13x9&11.99 $\pm$ 0.09 & 10.91 $\pm$ 0.06 & 10.69 $\pm$ 0.13 & 8.68 $\pm$ 0.08&	T \\
078&023.78286+0.50238 & Bp & 6x3 &  & &  &  &  & 7.10 $\pm$ 0.05 &	c \\
079&023.89021-0.73778 & Bs & 25x14 &  & &9.90 $\pm$ 0.02 & 8.32 $\pm$ 0.01 & 7.42 $\pm$ 0.01 & 5.56 $\pm$ 0.01&	P \\
080&023.90016-1.28024 & Brs/Ers & 18x10 & Ers & 12x9&  &  &  &  &	T \\
081&024.58540+0.11989 & Bs & 10x3 & S & &15.20 $\pm$ 0.82 &14.10 $\pm$ 0.54 & & &	p \\
082&024.76272-0.91396 & Bs & 3.4x2.2 &  & &14.22 $\pm$ 0.24 & 13.51 $\pm$ 0.20 & 12.89 $\pm$ 0.46 & 11.10 $\pm$ 0.29&	t \\
083&024.77483-1.31616 & Ars & 10 &  & &  &  &  &  &	c \\
084&024.89596+0.45853 & Bs & 14x6 &  & &10.69 $\pm$ 0.04 & 9.32 $\pm$ 0.02 & 8.31 $\pm$ 0.02 & 7.06 $\pm$ 0.02&	t \\
085&025.66408+1.15020 & As & 46 & As & 41x22 & & & &8.25 $\pm$ 0.08 &	P \\
086&025.77993-0.44005 & Rrs & 4.5x3.5 &  & &14.00 $\pm$ 0.22 & 12.79 $\pm$ 0.13 & 12.53 $\pm$ 0.40 & 10.37 $\pm$ 0.18&	p \\
087&025.92671-0.98449 & Bps & 13x5 & B & 4.6&12.07 $\pm$ 0.07 & 11.23 $\pm$ 0.05 & 10.51 $\pm$ 0.08 & 8.65 $\pm$ 0.05	&T \\
088&025.99096-0.59183 & Ers & 8x5 &  &  & & & &11.27 $\pm$ 0.39 &	c \\
089&026.42837+1.03759 & Bps & 7x6 &  & &13.63 $\pm$ 0.21 & 12.19 $\pm$ 0.09 & 12.05 $\pm$ 0.21 & 10.30 $\pm$ 0.18&	t \\
090&026.44767-0.80840 & Bps & 10x5 &  & & & & & &	c \\
091&026.74999-1.21865 & Ers & 18x12 & Ear & 15x13&  &  &  &  &	T \\
092&026.79572-1.05024 & Bs & 4x3 & E & 13x10&13.42 $\pm$ 0.16 & 11.78 $\pm$ 0.07 & 11.16 $\pm$ 0.10 & 9.28 $\pm$ 0.06	&T \\
093&026.83269-0.15180 & Bs & 9x3 & E & 7x5&11.55 $\pm$ 0.06 & 10.87 $\pm$ 0.06 & 10.36 $\pm$ 0.12 & 8.78 $\pm$ 0.09&	T \\
094&026.83640+0.28828 & Ers & 6x4 &  & & & & & &	c \\
095&027.09954+0.94886 & Es & 12x9 &  & & & & & & 	c \\
096&027.37280+1.39262 & Ear & 7.5x6.0 & R & 7 & & & & &	c \\
097&027.66357-0.82670 & Bps & 14x5.2 & Bmp & 35x12&12.37 $\pm$ 0.08 & 10.91 $\pm$ 0.04 & 9.68 $\pm$ 0.04 & 7.65 $\pm$ 0.02&	T \\
098&027.70327+0.70354 & Rars & 52x40 & Es & 15x12&9.76 $\pm$ 0.02 & 8.65 $\pm$ 0.01 & 8.53 $\pm$ 0.02 & 6.36 $\pm$ 0.01&	T \\
099&027.81843-0.76628 & Bs & 8x4 & S & &12.66 $\pm$ 0.11 & 11.97 $\pm$ 0.09 & 11.74 $\pm$ 0.17 & 9.80 $\pm$ 0.13&	T \\
100&028.06295-0.61048 & Rr & 6 &  &  & & & & &	c \\
101&028.19767-0.89109 & B & 14x8 &  &  & & & & &	c \\
102&028.52225-1.48422 & Rars & 30x30 & Bam & 44x42&  &  &  &  &	T \\
103&028.62122-0.86537 & Rr & 10 &  &  & & & & &	c \\
104&028.89451-0.29151 & Bs & 50x7 & As & 51x30&  &  &  &  &	T \\
105&029.21554+0.02262 & B & 30x12 &  &  & & & & &	c \\
106&029.50204+0.62395 & B & 25x12 & Bm & 24x13 &  &  &  &  &	T \\
107&029.57883-0.26901 & Bs & 6x3 & Sm & 4&9.36 $\pm$ 0.02 & 8.43 $\pm$ 0.01 & 7.51 $\pm$ 0.02 & 5.82 $\pm$ 0.01&	L \\
108&029.99765+0.65621 & B & 110x40 &  &  & & & & &	c \\
109&030.04497+0.03465 & Ers & 36x24 & Rar & 36&  &  &  &  &	T \\
110&030.17049+0.68782 & Ers & 5.4x3.2 &  &  & & & & &	c \\
111&030.22594+0.54285 & As & 15x15 &  & &9.21 $\pm$ 0.02 & 8.96 $\pm$ 0.02 & 9.07 $\pm$ 0.06 & 8.12 $\pm$ 0.07& 	YSO? \\
112&030.30097-1.22812 & Bs & 9x5 &  &  & & & & &	c \\
113&030.50743-0.21913 & Bs & 29x18 & Bs & 35x33&   &  &  &  &	T \\
114&030.66759-0.33136 & Bs & 15x10 &  & &9.46 $\pm$ 0.02 & 7.29 $\pm$ 0.01 & 5.97 $\pm$ 0.01 & 3.78 $\pm$ 0.00&	c \\
115&030.72160+0.14788 & Bs & 18x17 &  &  & & & & &	c \\
116&030.76828+1.40983 & B & 4x2.5 &  &  & & & & &	c \\
117&031.16908+0.81029 & R & 3.5 &  &  & & & & &	c \\
118&031.32618-0.53286 & Brs & 22x11 & Baprs & 34x28&13.07 $\pm$ 0.39 & 11.64 $\pm$ 0.20 & 10.73 $\pm$ 0.28 & 8.71 $\pm$ 0.15&	T \\
119&031.63781+0.99595 & Rr & 6 &  &  & & & & &	c \\
120&031.90685-0.30936 & Bs & 40x25 & Bmps & 69x42 &  &  &  &  &	T \\
121&032.14993+0.64445 & Bs & 9x5 &  &  & & & & &	c \\
122&032.22860-1.44045 & Es & 26x26 & Bmps & 47x45&  &  &  &  &	T \\
\end{tabular}
\end{table*}
\renewcommand{\arraystretch}{1.0}

\renewcommand{\arraystretch}{0.8}
\setcounter{table}{0}
\begin{table*}
\caption{continued.}
\begin{tabular}{llccccccccr}
\hline
No.&UWISH2 ID &  \multicolumn{2}{c}{\htwo} & \multicolumn{2}{c}{\ha} & \multicolumn{4}{c}{IRAC mag.}  & Type \\
   &       &    Morph. & Size           & Morph. & Size           & [3.6] & [4.5] & [5.8] & [8.0]  &  \\
\hline
123&032.28479-0.27816 & Bs & 12x8 &  & &12.73 $\pm$ 0.22 & 12.24 $\pm$ 0.20 & 11.75 $\pm$ 0.46 & 9.87 $\pm$ 0.28&	t \\
124&032.29224-0.74568 & As & 5.4x2.6 &  & &12.68 $\pm$ 0.10 & 12.25 $\pm$ 0.11 & 11.69 $\pm$ 0.20 & 12.08 $\pm$ 0.94& YSO? \\
125&032.37721-0.55490 & Bps & 14x7 & R & 5&13.45 $\pm$ 0.23 & 12.53 $\pm$ 0.14 & 11.43 $\pm$ 0.19 & 9.48 $\pm$ 0.12&	T \\
126&032.46866+0.28147 & Bs & 4x2 &  & &13.38 $\pm$ 0.18 & 12.46 $\pm$ 0.14 & 12.25 $\pm$ 0.36 & 10.88 $\pm$ 0.34&	t \\
127&032.54650-0.03210 & Bps & 28x14 & B & 15x11&  &  &  &  &	T \\
128&032.54998-0.29529 & Bps & 16x12 & E & 15x12&12.00 $\pm$ 0.11 & 11.75 $\pm$ 0.13 & 10.80 $\pm$ 0.19 & 8.93 $\pm$ 0.11&	T \\
129&032.61348+0.79678 & B & 5.6x4.0 & S & &13.38 $\pm$ 0.18 & 12.46 $\pm$ 0.14 & 12.25 $\pm$ 0.36 & 10.88 $\pm$ 0.34&	P \\
130&032.66916-1.25559 & Bps & 12x5 & E & 10x5&  &  &  & &	T \\
131&032.94004-0.74662 & Bs & 6x4 & E & 7x5&12.12 $\pm$ 0.07 & 11.67 $\pm$ 0.07 & 11.21 $\pm$ 0.13 & 9.03 $\pm$ 0.06&	T \\
132&033.16509+0.49150 & Bs & 6x3 &  & &13.45 $\pm$ 0.16 & 12.92 $\pm$ 0.14 & 11.40 $\pm$ 0.14 & 10.05 $\pm$ 0.16&	t \\
133&033.45470-0.61500 & As & 5x2.5 & S & 3.5& - &  &  &  &	T \\
134&033.88796+1.52134 & Bm & 38x20 & Bas & 51x37&  &  &  &  &	T \\
135&033.97946-0.98557 & Er & 10x8 & Ers & 8x7&  &  &  &  &	T \\
136&034.10462-1.64333 & Es & 10x7 & Es & 10x8&  &  &  &  &	T \\
137&034.41021+0.81477 & Es & 5x2 &  & &14.35 $\pm$ 0.30 & 13.27 $\pm$ 0.22 & 12.91 $\pm$ 0.50 & 11.18 $\pm$ 0.36&	t \\
138&034.84509+1.31721 & Bs & 14x7 &  &  &13.60 $\pm$ 0.16 &12.39 $\pm$ 0.10 & & &	L \\
139&035.18522+1.12134 & Ems & 12x9 &  &  &13.18 $\pm$ 0.18 &12.12 $\pm$ 0.11 & & &	p \\
140&035.23366-1.13623 & Ers & 60x50 &  & & & & & &	c \\
141&035.38919-1.17506 & Bs & 7.5x6 &  & & & & & &	c \\
142&035.47394-0.43716 & Bs & 25x15 & B & 16x10&8.48 $\pm$ 0.01 & 7.29 $\pm$ 0.01 & 5.71 $\pm$ 0.00 & 4.08 $\pm$ 0.00& t \\
143&035.76967-1.24531 & Bs & 26x12 &  &  & & & & & 	L \\
144&035.81426+1.48019 & As & 9x7 &  & &  &  &  &  &	c \\
145&035.81489-0.25181 & Bs & 25x20 &  &  & & & & &	P \\
146&035.89918-1.14425 & Ers & 7x6 &  &  & & & & &	c \\
147&036.05309-1.36593 & Bms & 300x90 & Bmps & 270x120&  &  &  &  &	T \\
148&036.43225-1.91396 & Bps & 14 & Bmp & 21&  &  &  &  &	Sym? \\
149&036.46081+0.80581 & Er & 14x9 &  &  & & & & &	c \\
150&036.48189+0.15610 & Bps & 12x5 &  & &12.91 $\pm$ 0.14 & 11.96 $\pm$ 0.10 & 10.93 $\pm$ 0.12 & 9.52 $\pm$ 0.13&	t \\
151&036.98479-0.20330 & Rrs & 12 &  & &13.40 $\pm$ 0.21 & 12.07 $\pm$ 0.12 & 11.16 $\pm$ 0.17 & 9.73 $\pm$ 0.15&	t \\
152&037.14125+0.30341 & E & 10x7 &  &  & & & & &	c \\
153&037.41544-0.19254 & Er & 6x5 &  & &13.03 $\pm$ 0.13 & 11.87 $\pm$ 0.09 & 10.99 $\pm$ 0.15 & 9.51 $\pm$ 0.11&	t \\
154&037.96134+0.45337 & Bp & 11x3.5 &  & &11.30 $\pm$ 0.04 & 9.79 $\pm$ 0.03 & 8.17 $\pm$ 0.02 & 6.21 $\pm$ 0.01&	L \\
155&038.14463-0.57489 & A & 10x8 &  & &  &  &  &  &	c \\
156&038.83959+0.87057 & E & 8x6 &  &  & & & & &	 c \\
157&039.16222+0.78375 & B & 12x6 &  & &11.29 $\pm$ 0.04 & 10.35 $\pm$ 0.03 & 8.99 $\pm$ 0.03 & 7.14 $\pm$ 0.02&	t \\
158&039.26101-0.55123 & Bs & 30x20 &  &  & & & & &	c \\
159&039.64158-0.36822 & Bs & 5x2.5 &  & &11.28 $\pm$ 0.04 & 10.30 $\pm$ 0.03 & 10.09 $\pm$ 0.07 & 8.99 $\pm$ 0.06&	t \\
160&040.03148-1.30313 & Ear & 30x20 & R? & 2.5 & & & & &	c \\
161&040.36950-0.47517 & Rrs & 35x32 & Rrs & 31&9.76 $\pm$ 0.02 & 8.65 $\pm$ 0.01 & 8.53 $\pm$ 0.02 & 6.36 $\pm$ 0.01&	T \\
162&040.47073+1.10067 & Bs & 12x4 &  &  &12.15 $\pm$ 0.07  &10.92 $\pm$ 0.05 & &7.54 $\pm$ 0.02 &	l \\
163&040.53948-0.76310 & Bs & 12x7 &  & &13.54 $\pm$ 0.21 & 11.94 $\pm$ 0.09 & 10.82 $\pm$ 0.12 & 9.93 $\pm$ 0.18&	t \\
164&040.96700-1.22601 & Ears & 80x50 &  &  & & & & &	c \\
165&041.27043-0.69797 & Bas & 30x15 & Bs & 18&12.12 $\pm$ 0.07 & 11.67 $\pm$ 0.07 & 11.21 $\pm$ 0.13 & 9.03 $\pm$ 0.06&	T \\
166&041.99634+0.10743 & B & 35x25 &  & & & & &8.45 $\pm$ 0.14 &	c \\
167&042.12631+0.45706 & Bs & 9x9 &  & &13.24 $\pm$ 0.18 & 12.21 $\pm$ 0.12 & 11.54 $\pm$ 0.23 & 9.84 $\pm$ 0.14&	t \\
168&042.97101-1.07103 & As & 7x5 &  & &  &  &  &  &	c \\
169&043.10420-1.70207 & Bs & 12x8 &  & & & & & &	c \\
170&043.25830+1.50423 & Es & 3.5x2.5 & As & 5x2 &15.23 $\pm$ 0.33 &14.31 $\pm$ 0.24  & & &	p \\
171&043.65562-0.82777 & Rs & 5 & S & &10.65 $\pm$ 0.03 & 10.18 $\pm$ 0.03 & 7.99 $\pm$ 0.01 & 6.31 $\pm$ 0.01& 	\hii \\
172&044.18877+1.56732 & Is & 140x120 & R & 3&  &  &  &  &	T \\
173&044.34714+0.08637 & Ear & 9x8 &  & &13.21 $\pm$ 0.15 & 12.59 $\pm$ 0.13 & 12.14 $\pm$ 0.36 & 10.06 $\pm$ 0.19&	t \\
174&044.73433+0.26046 & Bps & 25x12 & B & 15x9&11.09 $\pm$ 0.04 & 10.41 $\pm$ 0.04 & 9.77 $\pm$ 0.05 & 8.11 $\pm$ 0.05&	T \\
175&044.93245-0.01060 & Bms & 40x25 & Bamp & 35x20&  &  &  &  &	T \\
176&045.44425-1.57085 & Rr & 14 &  &  & & & & &	c \\
177&045.45878-0.49801 & Bas & 25x13 &  & & & & & &	c \\
178&045.95707+0.69049 & Is & 26x17 &  &  & & & & &	c \\
179&046.09523+1.36603 & Ers & 9x5 & B & 9x7 & & & & &	c \\
180&046.63335+1.31220 & As & 16x10 & B? & 7 &11.63 $\pm$ 0.04 &11.35 $\pm$ 0.05 & & &	p \\
181&046.93735-0.54973 & Br & 8x5 &  & &14.04 $\pm$ 0.23 & 13.35 $\pm$ 0.19 & 12.13 $\pm$ 0.25 & 11.04 $\pm$ 0.30&	t \\
182&047.18522+0.44999 & Brs & 25x14 & S? & & & &11.18 $\pm$ 0.08 & 9.16 $\pm$ 0.08 & 	p \\
183&047.44521+0.61199 & Es & 11x8 &  &  & & & & &	c \\
184&047.50612-0.36750 & Es & 12x8 &  & &13.99 $\pm$ 0.27 & 12.70 $\pm$ 0.15 & 11.34 $\pm$ 0.16 & 10.47 $\pm$ 0.22&	t \\
\end{tabular}
\end{table*}
\renewcommand{\arraystretch}{1.0}

\renewcommand{\arraystretch}{0.8}
\setcounter{table}{0}
\begin{table*}
\caption{continued.}
\begin{tabular}{llccccccccr}
\hline
No. &UWISH2 ID &  \multicolumn{2}{c}{\htwo} & \multicolumn{2}{c}{\ha} & \multicolumn{4}{c}{IRAC mag.}  & Type \\
    &      &    Morph. & Size           & Morph. & Size           & [3.6] & [4.5] & [5.8] & [8.0]  &  \\
\hline
185&047.52536+0.32144 & Ears & 7x5 &  &  & & & & &	c \\
186&047.61228+1.08168 & Ers & 13x11 & Ears & 16x12&  &  &  &  &	T \\
187&048.03497+0.12296 & Bas & 7x6 &  & & & & & &	c \\
188&048.24968-0.46947 & Bs & 10x8 &  & & &13.04 $\pm$ 0.22  & &11.26 $\pm$ 0.53 &	p \\
189&048.71570-0.28960 & Bs & 44x24 & B & 13x7&  &  &  &  &	T \\
190&048.99884+0.77703 & Rrs & 8x8 &  & &13.29 $\pm$ 0.13 & 12.24 $\pm$ 0.09 & 11.69 $\pm$ 0.14 & 9.93 $\pm$ 0.15&	t \\
191&049.28561+0.00740 & Brs & 24x16 & Ear & 18x13&  &  &  &  &	T \\
192&049.86763+1.06075 & As & 30x20 & A? & 25x17 & & & & &	c \\
193&049.91632-1.08675 & Es & 11x8 &  &  & & & & &	c \\
194&050.04556-0.79804 & Bs & 12x12 & & &  & 12.31 $\pm$ 0.10 & 11.69 $\pm$ 0.21 & 10.42 $\pm$ 0.20&	l \\
195&050.48027+0.70434 & Bs & 16x8 & B & 14x6 & 7.02 $\pm$ 0.00 & 6.03 $\pm$ 0.00 & 5.17 $\pm$ 0.00 & 4.56 $\pm$ 0.00&	Sym? \\
196&050.55559+0.04506 & Bs & 16x5 & S & &9.73 $\pm$ 0.02 & 8.84 $\pm$ 0.02 & 7.61 $\pm$ 0.01 & 5.78 $\pm$ 0.01& 	L \\
197&050.66579+1.33631 & B & 6x3.5 & E &6x5 &  &  &  &  &	T \\
198&050.66912+0.00673 & A & 22x26 & Ea & 23 &  &  &  &  & 	T \\
199&050.71285-0.17840 & B & 6x3 &  & &11.52 $\pm$ 0.04 & 10.58 $\pm$ 0.03 & 10.89 $\pm$ 0.10 & 8.86 $\pm$ 0.05&	t \\
200&050.78036+1.18512 & Ra & 50x20 & R? & 2.5 & & & & &	c \\
201&050.90936+1.05076 & As & 5 &  & & 8.21 $\pm$ 0.01 & 7.98 $\pm$ 0.01 &  & 4.91 $\pm$ 0.01& l \\
202&050.92866+0.43874 & Eas & 11x7 &  &  & & & & &	c \\
203&051.36290+0.87878 & Rars & 6 &  & &13.64 $\pm$ 0.15 & 12.59 $\pm$ 0.11 & 11.99 $\pm$ 0.16 & 10.52 $\pm$ 0.14&	t \\
204&051.50791+0.17037 & Er & 12x10 & Emrs & 13x9&9.57 $\pm$ 0.02 & 8.66 $\pm$ 0.01 & 7.63 $\pm$ 0.01 & 5.61 $\pm$ 0.01&	T \\
205&051.76939+1.36491 & Ears & 26 & Ears & 30x20&  &  &  &  &	T \\
206&051.83306+0.28374 & E & 54x30 & E & 20x16&10.58 $\pm$ 0.03 & 9.46 $\pm$ 0.02 & 9.02 $\pm$ 0.03 & 7.03 $\pm$ 0.02&	T \\
207&052.32654-0.13737 & Eas & 8x6 &  &  & & & & &	c \\
208&052.46943-0.90047 & Bs & 9x5 &  & &13.96 $\pm$ 0.20 & 12.79 $\pm$ 0.11 & 11.89 $\pm$ 0.17 & 9.98 $\pm$ 0.13&	t \\
209&052.70187-1.04355 & A & 25x12 &  & &  &  &  &  &	c \\
210&053.04316-0.06957 & Bas & 4x3 &  & &13.38 $\pm$ 0.20 & 13.27 $\pm$ 0.24 & 12.35 $\pm$ 0.38 & 11.13 $\pm$ 0.40&	t \\
211&053.36023-0.54988 & Ear & 12x6 &  &  & & & & &	c \\
212&054.29190-0.23778 & Brs & 30x10 &  & & & & & &	c \\
213&054.71154+0.41990 & Bs & 20x14 & Ba & 19x12&  &  &  &  &	T \\
214&055.50747-0.55729 & E & 20x14 & E & 17x13&8.95 $\pm$ 0.01 & 7.94 $\pm$ 0.01 & 6.83 $\pm$ 0.01 & 4.68 $\pm$ 0.00&	T \\
215&055.85017+1.44210 & I & 10x8 &  &  &10.02 $\pm$ 0.02 &9.53 $\pm$ 0.02  & & &	p \\
216&056.16673-0.41918 & Bas & 36x22 & Bas & 40x30?&  &  &  &  &	T \\
217&056.34479-1.53764 & Bas & 25x16 & B? & 28x16 & & & & &	c \\
218&056.42331-0.37341 & Bas & 18x12 & Ias & 31&12.79 $\pm$ 0.17 & 11.38 $\pm$ 0.07 & 11.49 $\pm$ 0.26 & 9.29 $\pm$ 0.14&	L \\
219&056.48535-0.09364 & Ear & 5x2.5 &  &  & & & & &	c \\
220&056.52321+0.30702 & Rr & 9 &  &  & & & & &	c \\
221&056.61303-0.04761 & E & 20x13 &  &  & & & & &	c \\
222&057.32913+0.61698 & Ear & 9x8 &  &  & & & & &	c \\
223&057.59474+0.50715 & Is & 70x40 &  & & & & & &	c \\
224&057.64365+0.47814 & B & 8x4 &  &  & & & & &	c \\
225&057.72078+0.12541 & Ear & 17x13 &  &  & & & & &	c \\
226&057.81415+0.78641 & Bs & 16x6 &  &  & & & & &	c \\
227&057.83552+1.04920 & Ears & 22x15 & R & 21 & & & &10.30 $\pm$ 0.31 &	P \\
228&057.98004-0.76740 & Brs & 26x20 & B & 27x30&12.12 $\pm$ 0.11 & 10.98 $\pm$ 0.06 & 10.02 $\pm$ 0.07 & 8.73 $\pm$ 0.08&	T \\
229&058.03770-0.04866 & Ir & 16x7 &  &  & & & & &	c \\
230&058.17873-0.81177 & Bs & 20x12 & B & 15x12&13.01 $\pm$ 0.18 & 12.19 $\pm$ 0.10 & 11.85 $\pm$ 0.29 & 10.38 $\pm$ 0.24&	L \\
231&058.80916+0.38692 & Bs & 20x20 &  &  & & & & &	c \\
232&059.18828-1.42144 & A & 20 & Ras & 27 & & & & &	L \\
233&059.36328+1.00137 & Er & 25x20 & R? & 25 & & & & &	c \\
234&059.77812-0.82788 & Bs & 12x9 & Ea & 13x9&9.41 $\pm$ 0.02 & 9.30 $\pm$ 0.02 & 9.08 $\pm$ 0.03 & 8.55 $\pm$ 0.04&	L \\
235&059.87554-0.60874 & Bas & 34x20 & Bas & 54x30&  &  &  &  &	T \\
236&060.24810+0.82261 & Bps & 35x14 & B & 7x6&12.64 $\pm$ 0.09 & 11.34 $\pm$ 0.05 & 10.72 $\pm$ 0.07 & 8.96 $\pm$ 0.05&	T \\
237&060.31487+0.79769 & Bs & 5x2.5 & S? & &14.15 $\pm$ 0.16 & 13.51 $\pm$ 0.14 & 11.59 $\pm$ 0.11 & 9.82 $\pm$ 0.07&	t \\
238&060.40130+0.97372 & A & 16x8 &  & &  &  &  &  &	c \\
239&060.52372-0.31828 & Bps & 11x6 & Bs & 11x6&12.62 $\pm$ 0.09 & 11.98 $\pm$ 0.08 & 10.97 $\pm$ 0.10 & 9.47 $\pm$ 0.07&	T \\
240&060.79926+1.17327 & R & 8 & R? & 5.5 & &14.79 $\pm$ 0.54 & & &	c \\
241&061.84215+0.88506 & Bs & 12x13 & B? & 12x12&12.92 $\pm$ 0.09 & 12.60 $\pm$ 0.11 & 12.26 $\pm$ 0.19 & 10.52 $\pm$ 0.19&	t \\
242&061.91270+0.20109 & Ear & 11x10 &  &  & & & & &	c \\
243&062.07780-0.43633 & Ear & 24x12 &  &  & & & & &	c \\
244&062.13719+0.14857 & Ear & 12x10 & E? & 10x8&12.96 $\pm$ 0.10 & 12.25 $\pm$ 0.10 & 11.49 $\pm$ 0.14 & 10.37 $\pm$ 0.15&	t \\
245&062.15368+1.15140 & E & 14x11 &  &  & & & & &	c \\
246&062.29042+1.13629 & Bs & 16x10 & E? & 11 & 12.83 $\pm$ 0.11 & 11.92 $\pm$ 0.07 &  & 9.80 $\pm$ 0.10 &	l \\
247&062.45283-0.01779 & Bs & 14x18 &  &  & & & & &	c \\
\end{tabular}
\end{table*}
\renewcommand{\arraystretch}{1.0}

\renewcommand{\arraystretch}{0.8}
\setcounter{table}{0}
\begin{table*}
\caption{continued.}
\begin{tabular}{llccccccccr}
\hline
No. &UWISH2 ID &  \multicolumn{2}{c}{\htwo} & \multicolumn{2}{c}{\ha} & \multicolumn{4}{c}{IRAC mag.}  & Type \\
    &      &    Morph. & Size           & Morph. & Size           & [3.6] & [4.5] & [5.8] & [8.0]  &  \\
\hline
248&062.49346-0.27008 & Bms & 70x18 & Bms & 90x25&11.87 $\pm$ 0.06 & 10.75 $\pm$ 0.04 & 10.15 $\pm$ 0.05 & 8.20 $\pm$ 0.04&	T \\
249&062.70165+0.06019 & Bs & 17x8 & Bps & 13x7&11.19 $\pm$ 0.04 & 10.68 $\pm$ 0.04 & 10.25 $\pm$ 0.06 & 8.82 $\pm$ 0.07&	T \\
250&062.75413-0.72565 & Bs & 80x10 & B & 30x10&12.06 $\pm$ 0.05 & 10.56 $\pm$ 0.03 & 9.60 $\pm$ 0.04 & 7.95 $\pm$ 0.03&	L \\
251&062.97476+1.38380 & Br & 15x9 & B & 14x8&  &  &  &  &	T \\
252&063.92454-1.21740 & Bs & 26x16 & Bamps & 100x80&  &  &  &  &	T \\
253&064.13697-0.97667 & Bprs & 20x20 & Bars & 33x17&  &  &  &  &	T \\
254&064.18792+0.77438 & Bs & 11x4.5 & E? & 2.3x4&13.28 $\pm$ 0.13 & 12.58 $\pm$ 0.11 & 11.32 $\pm$ 0.11 & 10.46 $\pm$ 0.15&	t \\
255&064.29941-0.14559 & Bas & 40x24 & Ears? & 23x20 & & & & &	c \\
256&064.94759+0.76048 & Brs & 10x8 & B? & 10x8.5 & & & &12.20 $\pm$ 0.39 &	c \\
257&065.54459+0.81855 & Br & 24x7 &  &  & & & & &	c \\
258&075.90338+0.29517 & Rr & 2.8 &  &  & & & & &	c \\
259&076.37264+1.17216 & Bprs & 44x30 & Bprs & 42x2&  &  &  &  &	T \\
260&076.88532+2.22199 & Bps & 7x4 &  &  &14.50 $\pm$ 0.38 &13.74 $\pm$ 0.30 & & &	p \\
261&077.65952-0.98321 & E & 20x17 &  &  & & & & &	c \\
262&077.68068+3.12797 & Bps & 40x6 & Bs & 22x5 &11.52 $\pm$ 0.05  & & & &	L \\
263&077.77375+1.55436 & R & 5 &  &  & & & & &	c \\
264&077.84010+0.86042 & Bs/Is & 20x15 &  & & & & & &	c \\
265&078.92993+0.76378 & Ras & 11x10 & Ras & 12x11&  &  &  &  &	T \\
266&079.33319+2.14863 & E & 8x5 &  &  & & & & &	c \\
267&079.62439+0.40225 & Bp & 26x12 &  & & & & & &	c \\
268&079.77014+1.89347 & R & 8 &  &  & & & & &	c \\
269&080.26214+0.24219 & Bs & 6x4 & E? & 7x6 & & & & &	c \\
270&081.70275+2.15524 & Ear & 12x5 &  &  & & & & &	c \\
271&082.02890-0.30589 & A & 15x10 &  & &  &  &  &  &	c \\
272&084.20031+1.09069 & Brs & 70x45 & Bmrs & 90x43&  &  &  &  &	T \\
273&084.68426-0.72166 & Rr & 9 &  &  & & & & &	c \\
274&143.50140-2.81706 & Brs & 90x40 & Brs & 170x90 & & & & &	\hii \\
275&144.15931-0.50100 & Ears & 24x20 & Rar & 21x25&  &  &  &  &	T \\
276&146.29327+0.54871 & R & 6 &  &  &10.31 $\pm$ 0.02  &9.92 $\pm$ 0.02  & & &	\hii? \\
277&149.16730-0.22038 & R & 12 &  & & & & & &	c \\
278&149.43257-2.19327 & Ers & 15x13 & E? & 17x13 & & & & &	c \\
279&151.30910-0.74888 & S & 3 &  & &11.63 $\pm$ 0.04 &10.48 $\pm$ 0.03 & & &	p \\
280&153.77044-1.40652 & Bps & 7x5 & Es & 6.7x6.0&  &  &  &  &	T \\
281&357.65660+0.26265 & Ear & 6x5 &  &  & & & & &	c \\
282&358.23394-1.18468 & Bms & 30x17 & B & 27x20&11.23 $\pm$ 0.07 & 10.16 $\pm$ 0.05 & 9.16 $\pm$ 0.06 & 7.82 $\pm$ 0.06&	T \\
283&358.25962-1.91267 & I & 6x5 & S? &  &8.24 $\pm$ 0.01 & 7.49 $\pm$ 0.01 & 7.02 $\pm$ 0.01 & 6.39 $\pm$ 0.01& 	Sym? \\
284&359.35683-0.98000 & Bps & 50x14 & Bmps & 60x25&7.23 $\pm$ 0.01 & 6.15 $\pm$ 0.00 & 4.58 $\pm$ 0.00 & 2.51 $\pm$ 0.00&	T \\
285&000.26373-0.19745 & Eas & 20x14 &  & & & & & &	c \\
286&006.33697-0.61480 & Ea & 7x5 &  &  & & & & &	c \\
287&009.20083-1.37007 & Bas & 50x20 &  &  & & & & &	c \\
288&010.03253+0.91443 & Ba & 8x5 &  &  & & & & &	c \\
289&011.32349-1.66821 & Ra & 26 &  &  & & & & &	c \\
290&015.46686+1.10391 & B & 5x3 &  &  & & & & &	c \\
291&018.86276+0.65375 & B & 9x4 &  &  & & & & &	c \\
292&019.77912-0.85302 & E & 8x4 &  &  & & & & &	c \\
293&022.28851-0.06156 & B & 7x5 &  &  & & & & &	c \\
294&027.85065-0.64690 & E & 12x8 &  &  & & & & &	c \\
295&030.25432+1.53745 & B & 5x3 &Ers  &7x5 &13.22 $\pm$ 0.13 & 11.73 $\pm$ 0.06 & 11.42 $\pm$ 0.09 & 10.12 $\pm$ 0.11&	L \\
296&030.84096+0.07624 & Es & 25x20 &  &  & & & & &	c \\
297&032.37928-0.51843 & Ba & 8x3 &  &  & & & & &	c \\
298&033.60731+0.94113 & Bas & 17x12 &  & &12.90 $\pm$ 0.14 & 12.10 $\pm$ 0.10 & 11.29 $\pm$ 0.14 & 10.01 $\pm$ 0.17&	t \\
299&035.08538-0.53884 & Es & 4.5x3 &  &  & & & & &	c \\
300&044.60870-0.36523 & B & 4.5x3 &  &  & & & & &	c \\
301&044.67282-0.98346 & Bp & 17x6 &  &  & & & & &	c \\
302&051.75061+0.05155 & B & 16x6 &  &  & & & & &	c \\
303&060.02622-0.28202 & Bps & 7x3.5 &  &  & & & & &	c \\
304&063.19875+1.14092 & Bs & 18x8 &  &  & & & & &	c \\
305&064.12864+0.14814 & Bs & 15x6 &  &  & & & & &	c \\
306&078.65278+3.53770 & E & 26x15 &  &  & & & & &	c \\
307&081.12354+1.23547 & Bas & 19x11 &  &  & & & & &	c \\
\hline
\end{tabular}
\end{table*}
\renewcommand{\arraystretch}{1.0}


\renewcommand{\arraystretch}{0.8}
\begin{table*}
\caption{\htwo\ features conisdered possible PNe and which are
  additional to those in Table B1 of F15. Columns are (1) the UWISH2
  source ID with the Galactic coordinates, (2) decimal RA and (3)
  decimal Dec of the geometric centre of the \htwo\ feature, (4)
  radius enclosing the \htwo\ emission, (5) area covered by
  \htwo\ emission, (6) total and (7) median \htwo\ flux, (8) other
  identifier or comment.}
\begin{tabular}{lccccccl}
\hline
UWISH2 ID      &  RA    & DEC   & Radius  & Area  & $F_{\rm tot}$   & $F_{\rm med}$ & Other ID \\
               & [deg]  & [deg] & [arcsec] & [arcsec$^{2}$] & \multicolumn{2}{c}{[$10^{-19}$~W~m$^{-2}$]} & \\ 
\hline
000.26373-0.19745 & 266.75411 & -28.81354 & 14.8 & 253.29 & 2781.84 & 1441.22 \\
006.33697-0.61480 & 270.58274 & -23.78468 & 6.4  & 38.86  & 171.51  &167.487 \\
009.21329-1.36581 & 272.81301 &-21.63926 &49.5 &330.06 &1236.46 &1218.34 \\
010.03253+0.91443  &271.10781  &-19.81514 &9.1 &72.41 &270.59 &255.61 \\
011.32349-1.66821  &274.17547  &-19.93079 &10.1 &71.48 &238.36 &217.38\\
015.46686+1.10391 &273.68854 &-14.96764 &3.7 &26.00 &119.46 &125.58\\
018.86358+0.65383 &275.74921 &-12.18806 &6.9 &20.90 &61.61 &64.28\\
019.77912-0.85302 &277.54931 &-12.08032 &5.5 &16.31 &21.71 &3.54&\\
022.28851-0.06156 &278.01694 &-9.48958 &5.1 &31.28 &109.98 &108.23\\
027.85065-0.64690 &281.11497 &-4.81676 &8.7 &72.76 &292.47 &264.85\\
030.25432+1.53745 &280.26926 &-1.68065 &3.2 &15.43 &57.72 &52.31&PN\,G030.2+01.5\\
030.84096+0.07624 &281.83781 &-1.82613 &10.8 &76.02 &372.25 &207.535\\
032.37928-0.51843 &283.06890 &-0.72829 &5.7 &53.88 &274.83 & 7252.70\\
033.60731+0.94113 &282.32974 &+1.03010 &12.0 &201.31 &1293.18 &841.476\\
035.08538-0.53884 &284.32148 &+1.67049 &2.7 &13.71 &56.61 &56.3481\\
044.60870-0.36523 &288.57179 &+10.20750 &3.5 &16.01 &44.58 &40.67\\
044.67282-0.98346 &289.15792 &+9.97660 &4.9 &26.42 &68.82 &63.67\\
051.75061+0.05155 &291.64503 &+16.70811 &6.0 &34.56 &136.74 &126.14\\
060.02622-0.28202 &296.25780 &+23.77211 &8.0 &58.69 &183.26 &168.44\\
063.19875+1.14092 &296.65712 &+27.23026 &3.9 &20.48 &48.24 &44.65\\
064.12864+0.14814 &298.15198 &+27.52725 &7.0 &51.06 &126.79 &111.82\\
078.65278+3.53770 &304.10530 &+41.61373 &12.9 &122.8 &343.84 &369.86\\
081.12354+1.23547 &308.54931 &+42.30322 &11.2 &161.89 &451.45 &422.53\\
\hline
\end{tabular}
\end{table*}
\renewcommand{\arraystretch}{1.0}
\twocolumn



\begin{thebibliography}{99}

\bibitem[\protect\citeauthoryear{Aleman}{2011}]{aleman11}
Aleman I., Gruenwald R., 2011, A\&A, 528, 74

\bibitem[\protect\citeauthoryear{Barentsen}{2014}]{barentsen14}
Barentsen G., et al., 2014, MNRAS, 444, 3230

\bibitem[\protect\citeauthoryear{Becker}{1994}]{becker94}
Becker R.H., White R.L., Helfand D.J., Zoonematkermani S., 1994, ApJS, 91, 347

\bibitem[\protect\citeauthoryear{Beckwith}{1978}]{beckwith78}
Beckwith S., Persson S.E., Gatley I., 1978, ApJ, 219, 33

\bibitem[\protect\citeauthoryear{Casali}{1996}]{casali96}
Casali M.M., Eiroa C., 1996, A\&A, 306, 427

\bibitem[\protect\citeauthoryear{Choi}{2016}]{choi16}
Choi J., Dotter, A., Conroy C., Cantiello M., Paxton B., Johnson B.D., 2016, ApJ, 823, 102

\bibitem[\protect\citeauthoryear{Cohen}{2011}]{cohen11}
Cohen M., Parker Q.A., Green A.J., Miszalski B., Frew D., Murphy T., 2011, MNRAS, 413, 514

\bibitem[\protect\citeauthoryear{Cooper}{2013}]{cooper3}
Cooper H.D.B., et al., 2013, MNRAS, 430, 1125

\bibitem[\protect\citeauthoryear{Davis}{2003}]{davis03}
Davis C.J., Smith M.D., Stern L., Kerr T.H., Chiar J.E., 2003, MNRAS, 344, 262

\bibitem[\protect\citeauthoryear{Deharveng}{2010}]{deh10}
Deharveng L., et al., 2010, A\&A, 523, 6

\bibitem[\protect\citeauthoryear{Drew}{2005}]{drew05}
Drew J.E., et al., 2005, MNRAS, 362, 753

\bibitem[\protect\citeauthoryear{Drew}{2014}]{drew14}
Drew J.E., et al., 2014, MNRAS, 440, 2036

\bibitem[\protect\citeauthoryear{Lucas}{2008}]{lucas08}
Lucas P.W., et al., 2008, MNRAS, 391, 136 

\bibitem[\protect\citeauthoryear{Fang}{2018}]{fang18}
Fang X., Zhang Y., Kwok S., Hsia C-H., Chau W., Ramos-Larios G., Guerrero M., 2018, arXiv:1804.08840

\bibitem[\protect\citeauthoryear{Flagey}{2011}]{flagey11}
Flagey N., Noriega-Crespo A., Billot N., Carey S.J., 2011, ApJ, 741, 4

\bibitem[\protect\citeauthoryear{Frew}{2016}]{frew14}
Frew D.J., Boji\v{c}i\'{c} I.S., Parker Q.A., Pierce M.J., Gunawardhana M.L.P.,
Reid W.A., 2014, MNRAS, 440, 1080

\bibitem[\protect\citeauthoryear{Frew}{2010}]{frew10}
Frew D.J., Parker Q.A., 2010, PASA, 27. 129

\bibitem[\protect\citeauthoryear{Frew}{2016}]{frew16}
Frew D.J., Parker Q.A., Boji\v{c}i\'{c} I.S., 2016, MNRAS, 455, 1459

\bibitem[\protect\citeauthoryear{Froebrich}{2011}]{froebrich11}
Froebrich D., et al., 2011, MNRAS, 413, 480

\bibitem[\protect\citeauthoryear{Froebrich}{2015}]{froebrich15}
Froebrich D., et al., 2015, MNRAS, 454, 2586

\bibitem[\protect\citeauthoryear{Gledhill}{2015}]{gled15}
Gledhill T.M., Forde K.P., 2015, MNRAS, 447, 1080

\bibitem[\protect\citeauthoryear{Guerrero}{2000}]{guerrero00}
Guerrero M.A., Villaver E., Manchado A., Garcia-Lario P., Prada F., 2000, ApJS, 127, 125 

\bibitem[\protect\citeauthoryear{Hrivnak}{1994}]{hrivnak94}
Hrivnak B.J., Kwok S., Geballe T.R., 1994, ApJ, 420, 783

\bibitem[\protect\citeauthoryear{Jacob}{2013}]{jacob13}
Jacob R., Sch\"{o}nberner D., Steffen M., 2013, A\&A, 558, 78

\bibitem[\protect\citeauthoryear{Kanarek}{2015}]{kanarek15}
Kanarek G., Shara M., Faherty J., Zurek D., Moffat A., 2015, MNRAS, 452, 2858

\bibitem[\protect\citeauthoryear{Kastner}{1996}]{kastner96}
Kastner J.H., Weintraub D.A., Gatley I., Merrill K.M., Probst R.G., 1996, ApJ, 462, 777

\bibitem[\protect\citeauthoryear{Kharchenko}{2013}]{khar13}
Kharchenko N.V., Piskunov A.E., Schilbach E., R\"{o}ser S., Scholz R.-D., 2013, A\&A, 558, 53

\bibitem[\protect\citeauthoryear{Manchado}{1996}]{manch96}
Manchado A., Guerrero M.A., Stanghellini L., Serra-Ricart M., 1996, The IAC Morphological
Catalog of Northern Galactic Planetary Nebulae. Instituto de Astrofisica de Canarias (IAC),
La Laguna, Spain 

\bibitem[\protect\citeauthoryear{Marquez-Lugo}{2015}]{marquez-lugo15}
Marquez-Lugo R.A., Guerrero M.A., Ramos-Lario G., Miranda L.F.2015, MNRAS, 453, 1888 

\bibitem[\protect\citeauthoryear{Megeath}{2012}]{meg12}
Megeath S.T., et al., 2012, AJ, 144, 192

\bibitem[\protect\citeauthoryear{Misz}{2008}]{misz08}
Miszalski B., Parker Q.A., Acker A., Birkby J.L., Frew D.J., Kovacevic A., 2008, MNRAS, 384, 525

\bibitem[\protect\citeauthoryear{Mizuno}{2010}]{mizuno2010}
Mizuno D.R., et al., 2010, AJ, 139, 1542 

\bibitem[\protect\citeauthoryear{Nowak}{2014}]{nowak14}
Nowak M., Flagey N., Noriega-Crespo A., Billot N., Carey S.J., Paladini R., Van Dyk, S.D., 2014, ApJ, 796, 116

\bibitem[\protect\citeauthoryear{Ortega}{2013}]{ortega13}
Ortega M. E., Paron S., Giacani E., Rubio M., Dubner G., 2013, A\&A, 556, 105 

\bibitem[\protect\citeauthoryear{Parker}{2005}]{parker05}
Parker Q.A., et al., 2005, MNRAS, 362, 689

\bibitem[\protect\citeauthoryear{Parker}{2006}]{parker06}
Parker Q.A., et al., 2006, MNRAS, 373, 79

\bibitem[\protect\citeauthoryear{Parker}{2012}]{parker12}
Parker Q.A., et al., 2012, MNRAS, 427, 3016

\bibitem[\protect\citeauthoryear{Parker}{2016}]{parker16}
Parker Q.A., Boji\v{c}i\'{c} I.S., Frew D.J., 2016, in Liu X.,
Stanghellini L., Karakas A., eds, Proc. IAU Symp. 323, Planetary Nebulae:Multi-wavelength Probes 
of Stellar and Galactic Evolution. Cambridge University Press, p.36

\bibitem[\protect\citeauthoryear{Purcell}{2013}]{purcell13}
Purcell C.R., et al., 2013, ApJS, 205, 1

\bibitem[\protect\citeauthoryear{Ramos-Larios}{2017}]{rl17}
Ramos-Larios G., Guerrero M.A., Sabin L., Santamar\'{i}a E., 2017, MNRAS, 470, 3707

\bibitem[\protect\citeauthoryear{Reipurth}{1997}]{reipurth97}
Reipurth B., Aspin C., 1997, AJ, 114, 2700

\bibitem[\protect\citeauthoryear{Rowles}{2009}]{rowles09}
Rowles J., Froebrich D., 2009, MNRAS, 395, 164

\bibitem[\protect\citeauthoryear{Sabin}{2014}]{sabin14}
Sabin L., et al., 2014, MNRAS, 443, 3388

\bibitem[\protect\citeauthoryear{Storey}{1984}]{storey84}
Storey J.W.V., 1984, MNRAS, 206, 521

\bibitem[\protect\citeauthoryear{Treffers}{1976}]{treffers76}
Treffers R.R., Fink U., Larson H.P., Gautier T.N., 1976, ApJ, 209, 793

\bibitem[\protect\citeauthoryear{Urquhart}{2009}]{urq09}
Urquhart J.S., et al., 2009, A\&A, 501, 539

\bibitem[\protect\citeauthoryear{Wang}{2012}]{wang14}
Wang S., Jiang B.W., 2014, ApJ, 788, L12

\bibitem[\protect\citeauthoryear{Zhang}{2012}]{zhang12}
Zhang Y., Hsia C-H., Kwok S., 2012, ApJ, 745, 59

\bibitem[\protect\citeauthoryear{Zuckerman}{1988}]{zuck88}
Zuckerman B., Gatley I., 1988, ApJ, 324, 501
\end{thebibliography}
\end{document}